\documentclass[prb,twocolumn,showpacs,superscriptaddress, amsmath,amssymb]{revtex4}

\usepackage{graphicx}% Include figure files
\usepackage{dcolumn}% Align table columns on decimal point
\usepackage{bm}% bold math
\usepackage[dvips]{color}
\usepackage{color}

\newcommand{\vect}[1]{\mathbf{#1}}

\begin{document}
\preprint{APS/123-QED}

\title{Exotic phases of frustrated antiferromagnet LiCu$_2$O$_2$}

\author{A. A. Bush}
\affiliation{Moscow Institute of Radioengineering, Electronics and Automation, 117454 Moscow, Russia}

\author{N. B\"uttgen}
\affiliation{Center for Electronic Correlations and Magnetism EKM,
Experimentalphysik V, Universit\"at Augsburg, D–86135 Augsburg, Germany}

\author{A. A. Gippius}
\affiliation{Faculty of Physics, M. V. Lomonosov Moscow State University, 119991 Moscow, Russia}
\affiliation{P. N. Lebedev Physical Institute RAS, 119991 Moscow, Russia}

\author{M. Horvati\'c}
\affiliation{Laboratoire National des Champs Magn\'etiques Intenses, LNCMI-CNRS (UPR3228), \\ EMFL, UGA, UPS, and INSA, Bo\^{i}te Postale 166, 38042, Grenoble Cedex 9, France}

\author{M. Jeong}
\altaffiliation[Present address: ]{School of Physics and Astronomy, University of Birmingham, Birmingham B15 2TT,
United Kingdom}
\affiliation{Laboratoire National des Champs Magn\'etiques Intenses, LNCMI-CNRS (UPR3228), \\ EMFL, UGA, UPS, and INSA, Bo\^{i}te Postale 166, 38042, Grenoble Cedex 9, France}

\author{W.~Kraetschmer}
\affiliation{Center for Electronic Correlations and Magnetism EKM,
Experimentalphysik V, Universit\"at Augsburg, D–86135 Augsburg, Germany}

\author{V. I. Marchenko}
\affiliation{P. L. Kapitza Institute for Physical Problems RAS, 119334 Moscow, Russia}

\author{Yu. A. Sakhratov}
\affiliation{Kazan State Power Engineering University, 420066 Kazan, Russia}

\author{L. E. Svistov}
\email{svistov@kapitza.ras.ru}
\affiliation{P. L. Kapitza Institute for Physical Problems RAS, 119334 Moscow, Russia}

\date{\today}

\begin{abstract}
$^{7}$Li NMR spectra were measured in a magnetic field up to 17~T at temperatures 5-30~K on single crystalline LiCu$_2$O$_2$. Earlier reported anomalies on magnetization curves correspond to magnetic field values where we observe changes of the NMR spectral shape. For the interpretation of the field and temperature evolutions of our NMR spectra, the magnetic structures were analyzed in the frame of the phenomenological theoretical approach of the Dzyaloshinskii-Landau theory. A set of possible planar and collinear structures was obtained. Most of these structures have an unusual configuration; they are characterized by a two-component order parameter and their magnetic moments vary harmonically not only in direction, but also in size. From the modeling of the observed spectra, a possible scenario of magnetic structure transformations is obtained.
\end{abstract}


\maketitle

\section{Introduction}

Unconventional magnetic orders and phases in frustrated quantum-spin chains appear under a fine balance of the exchange interactions and are sometimes caused by much weaker interactions or fluctuations.\cite{Chubukov_1991, Kolezhuk_2000, Kolezhuk_2005, Dmitriev_2008, Amiri_2015} A kind of frustration in quasi-one-dimensional (1D) chain magnets is provided by competing interactions, when the intrachain nearest neighbor (NN) exchange is ferromagnetic ($J_{NN} < 0$) and the next-nearest neighbor (NNN) exchange is antiferromagnetic ($J_{NNN} > 0$). Numerical investigations of frustrated chain magnets within different models\cite{Hikihara_2008, Sudan_2009, Kolezhuk_2009} have predicted a number of exotic magnetic phases in the magnetization process, such as planar, spiral and different multipolar phases. The theoretical study of the magnetic phase diagram shows that real magnetic phases are very sensitive to interchain interactions and anisotropic interactions.

LiCu$_2$O$_2$ is an example of Cu$^{2+}$ ($S = 1/2$) magnet with frustrated
exchange interactions akin to the quantum spin-chain compound LiCuVO$_4$.\cite{Prozorova_2015, Buttgen_2014} However, the magnetic structure of LiCu$_2$O$_2$ appears to be complicated by interchain interactions between coupled
chains of magnetic Cu$^{2+}$ ions. Superexchange interactions via oxygen ions of edge-shared CuO$_4$ squares (see Fig.~1) provide frustration of the intrachain exchange interactions ($J_{NN} < 0$, $J_{NNN} > 0$).\cite{Masuda_2005}
According to Ref.~[\onlinecite{Hikihara_2008}], for the 1D model with the intrachain exchange constants of LiCu$_2$O$_2$, a chiral long-range order in the low-field range and a quasi-long-range ordered spin-density-wave phase in higher applied magnetic fields $H$ are expected similar to the above mentioned LiCuVO$_4$. Experimentally an incommensurate spin structure was observed at $T < T_N$  in the low-field range, which was ascribed to a planar helical spin structure.\cite{Masuda_2004, Gippius_2004}
Despite the fact that the magnetic properties of LiCu$_2$O$_2$ and structurally isomorphic NaCu$_2$O$_2$ have been intensively studied for more than ten years, the magnetic structures of these compounds have not been unambiguously determined. The lack of a reliable interpretation of the magnetic structure makes it impossible to unequivocally explain the nature of the multiferroic properties of LiCu$_2$O$_2$ and their absence in NaCu$_2$O$_2$.\cite{Leininger_2010}

In the present work,  $^7$Li NMR spectra of untwinned single crystals of LiCu$_2$O$_2$ were studied in a magnetic field up to 17~T at temperatures 5-30~K. Field and temperature dependencies of the spectra are discussed in Sec.~IV. The evolving magnetic structures were analyzed in the frame of the Dzyaloshinskii-Landau  theory of magnetic phase transitions. This analysis was performed in the exchange approximation, i.e., under the assumption that the exchange interactions predominate the interactions of relativistic nature (Sec.~V).
A theoretical analysis of the relativistic effects, such as anisotropy and electric polarization, is given in Appendix~B. From the simulations of the NMR line shape of our observed spectra we elaborate
the underlying magnetic structures theoretically and provide a scenario for the case of LiCu$_2$O$_2$. The structures exhibit an extraordinary configuration; a two-component order parameter characterizes the ordering of the magnetic moments which appear to be not only rotated  but also harmonically modulated in size from site to site (Sec.~VI).
Note that elliptical structures in magnets with a two-component order parameter arise quite often.\cite{Frontzek_2012, Willenberg_2012, Willenberg_2016, Khasanov_2017, Matsuda_2007, Rodriguez_2011, Ye_2012} However, it is commonly accepted that the ellipticity of the spiral structure is due to relativistic interactions, while the proposed structures for LiCu$_2$O$_2$ are elliptical already in the exchange approximation.

\section{Crystal and magnetic structure}

LiCu$_2$O$_2$ crystallizes in an orthorhombic lattice (space group $Pnma$)
with the unit cell parameters $a=$5.73 \AA, $b=$2.86 \AA \, and $c=$12.42
\AA.\cite{Berg_1991} The unit cell parameter $a$ is approximately twice the
unit cell parameter $b$. Consequently, LiCu$_2$O$_2$ single crystals, as a rule, exhibit considerable
twinning due to the formation of crystallographic domains rotated by $90^\circ$ around their crystallographically common $c$ axis.

The unit cell of the LiCu$_2$O$_2$ crystal contains
four monovalent nonmagnetic cations Cu$^+$ and four divalent cations Cu$^{2+}$
with $S$=1/2. The positions of all ions in the crystal lattice are schematically shown in Fig.~1.
The unit cell is selected with a dashed line. There are four crystallographically equivalent positions of
the magnetic Cu$^{2+}$ ions in the crystal unit cell of LiCu$_2$O$_2$,
denoted as I, II, III, and IV.

\begin{figure}
\includegraphics[width=0.7\columnwidth,angle=0,clip]{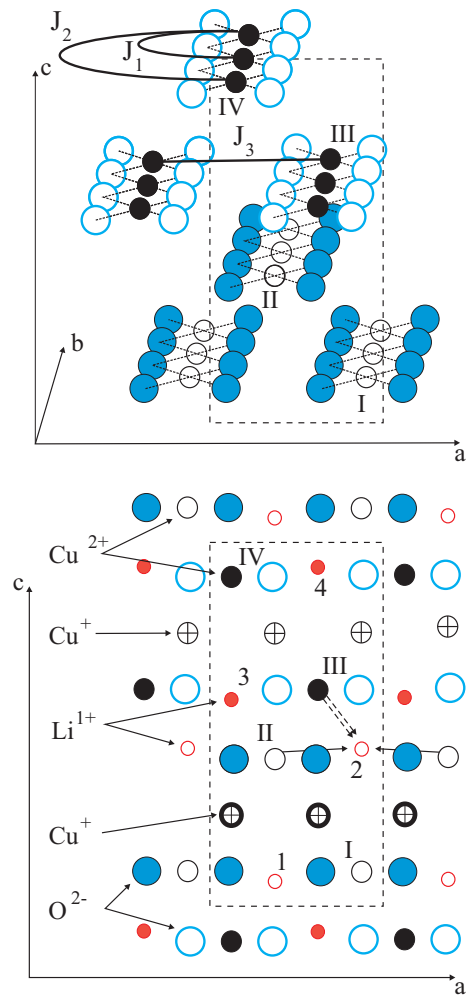}
\caption{(color online)
Bottom: The crystal structure of LiCu$_2$O$_2$ projected onto the $ac$-plane, all ions are shown.
The positions of the magnetic Cu$^{2+}$ ions are marked I, II, III and IV. The positions of Li$^+$ ions
are marked 1, 2, 3 and 4. The two arrows from II to 2 and the two arrows from III to 2 show possible hyperfine links
discussed in Sec.~VI.
Top: Cu$^{2+}$ magnetic chains and nearest O$^{2-}$ ions are shown.
}
\label{fig:fig1}
\end{figure}

The two-stage transition into a magnetically ordered state occurs at
$T_{c1}=24.6$~K and $T_{c2}=23.2$~K.\cite{Seki_2008} Neutron scattering and NMR experiments
revealed an incommensurate magnetic structure in the magnetically ordered state
($T<T_{c1}$).\cite{Masuda_2004, Gippius_2004, Kobayashi_2009} The wave vector
of the incommensurate magnetic structure coincides with the chain direction
($b$-axis). The magnitude of the propagation vector at $T$ $<$ 17 K is almost
temperature independent and is equal to 0.827$\times$2$\pi$/$b$. The neutron
scattering experiments have shown that that adjacent magnetic moments along the $a$-direction are oriented antiparallel, whereas those along the $c$-direction are mutually coaligned.
The intra-chain and inter-chain exchange constants were determined from the analysis of the spin wave
spectra.\cite{Masuda_2005} The large number of exchange bonds stipulates the ambiguity of the main exchange parameters obtained from Ref.~[\onlinecite{Masuda_2005}]. Theoretical analysis based on local density approach
(LDA) and cluster calculations as well as a phenomenological analysis of the temperature dependence of the magnetic susceptibility allow to map the most significant exchange paths. Using the result of these investigations\cite{Drechsler_2005, Maurice_2012, Mazurenko_2007} we choose the most suitable set of parameters proposed from neutron experiments: the nearest neighbor exchange interaction is ferromagnetic $J_1=-7.00$~meV, while the next nearest neighbor exchange interaction is antiferromagnetic $J_2=3.75$~meV.
The competition between these intra-chain interactions leads to an incommensurate magnetic structure.
The antiparallel orientation of magnetic moments of Cu$^{2+}$ between nearby chains is
caused by strong antiferromagnetic exchange interaction $J_3=3.4 $~meV.
These main exchange paths are shown in Fig.~1.
The coupling of the Cu$^{2+}$ moments along the $c$ direction and the couplings
between the magnetic ions in other crystallographic positions are much
weaker.\cite{Masuda_2005, Gippius_2004} Thus, LiCu$_2$O$_2$ can be considered
as a quasi-two dimensional system. The quasi-two-dimensional character of magnetic
interactions in LiCu$_2$O$_2$ compound was also proved by resonant soft
x-ray magnetic scattering experiments.\cite{Rusydi_2008, Huang_2008}

The magnetic structure of LiCu$_2$O$_2$ at zero magnetic field was studied by
several groups by means of neutron diffraction experiments.\cite{Masuda_2004,
Seki_2008, Kobayashi_2009} The authors of Ref.~[\onlinecite{Masuda_2004}] have
proposed the planar spiral spin structure with the spins confined to the
$ab$ plane. Polarized neutron scattering measurements\cite{Seki_2008} have detected the spin component along the
$c$ direction, indicating the spiral magnetic structure in the $bc$ plane.
The authors of Ref.~[\onlinecite{Kobayashi_2009}], alternatively, have proposed a
spiral spin structure confined to the (1,1,0) plane. It was attempted to
extract information about the magnetic structure of LiCu$_2$O$_2$ from the studies
of electric polarization, which accompanies magnetic
ordering.\cite{Park_2007,Seki_2008} Unfortunately, at the moment, the nature
of this polarization  is not clear\cite{Moskvin_2009} and, hence,
does not allow to draw an unambiguous conclusion about the zero-field
magnetic structure from this type of experiment.

The magnetic structure of LiCu$_2$O$_2$ was also studied by $^{63,65}$Cu and $^{7}$Li NMR in Ref.~[\onlinecite{Sadykov_2012}]. The authors of this work describe their results in the frame of planar spin structure and come to the conclusion that the spiral planes do not coincide with any of the crystallographic planes $ab$, $ac$, or $bc$, respectively.

\section{Sample preparation and experimental details}

Untwinned single crystals of LiCu$_2$O$_2$ with the size of several cubic
millimeters were prepared by the solution in the melt method as described in Ref.~[\onlinecite{Svistov_2009}].
The quality of the crystals under investigation was studied in Refs.~[\onlinecite{Kamentsev_2013, Bush_2004}] and the magnetic properties were found to be identical for all the samples from different batches.

$^{7}$Li nuclei (nuclear spin $I=3/2$, gyromagnetic ratio $\gamma/2\pi=16.5471$~MHz/T) were
probed using pulsed NMR technique. The spectra were obtained by summing fast Fourier transforms (FFT)
or integrating the averaged spin-echo signals as the field was swept through the resonance line.
NMR spin echoes were obtained using $\tau_p~-~\tau_D~-~\tau_p$ pulse sequences, where the pulse
lengths $\tau_p$ were 1.5~$\mu$s, the delay times between the pulses $\tau_D$ were 28~$\mu$s.
Measurements were carried out in the temperature range $5\leq T \leq 30$~K stabilized with a
precision better than 0.1~K.

\section{Experimental results}

$^7$Li NMR spectra were studied for four orientations of the static magnetic field:
$\vect{H}\parallel \vect{a}$ (Figs.~2, 3, 4), $\vect{H}\parallel \vect{b}$ (Figs.~5 and 10),
$\vect{H}\parallel \vect{c}$ (Figs.~6, 11, and 12), and $\vect{H}\parallel (\vect{c}+15^\circ)$ ($\vect{H}$ in $bc$ plane at an angle of 15$^\circ$ with respect to $c$ axis, Fig.~7).

\begin{figure}
\includegraphics[width=0.95\columnwidth,angle=0,clip]{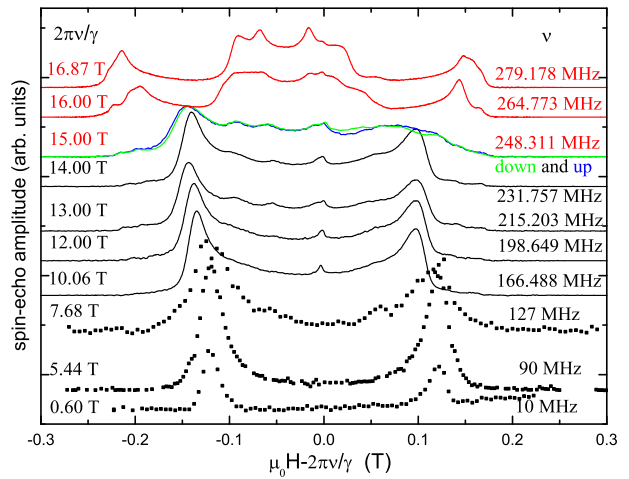}
\caption{(color online)
Field evolution of $^7$Li NMR spectra, $\vect{H}\parallel \vect{a}$, $T = 5$~K (4.5~K for the frequencies 10, 90, 127~MHz, taken from Ref.~[\onlinecite{Svistov_2009}]). Low-field phase (black) and high-field phase (colored).
}
\label{fig:fig2}
\end{figure}

Figures~2, 5, 6, and 7 show field evolutions of the spectra. The measurements were performed at the lowest temperature 5~K (4.5~K for $\vect{H}\parallel \vect{a}$ and $\nu = 10, 90, 127$~MHz), i.e., well below the magnetic ordering temperature ($T_N \approx 24$~K). The spectra lines are shifted along the vertical axis for clarity. The resonance field is defined by the vector sum of the external field $\vect{H}$ and the effective field $\vect{H}_{eff}$ from neighboring magnetic environment. For the case $H \gg H_{eff}$ the resonance field $H_{res} \approx H + H_{eff}$, where $H_{eff}$ is the projection of the effective field on the direction of the external field. In figs.~2-7, the field scale for each NMR spectrum is shifted by the value of the undisturbed Larmor field $2\pi\nu / \gamma$ at $^7$Li nuclei ($\gamma / 2\pi = 16.5471$~MHz/T). In such a representation, the horizontal axes show $H_{eff}$ at the nuclei of nonmagnetic ions.

\begin{figure}
\includegraphics[width=0.95\columnwidth,angle=0,clip]{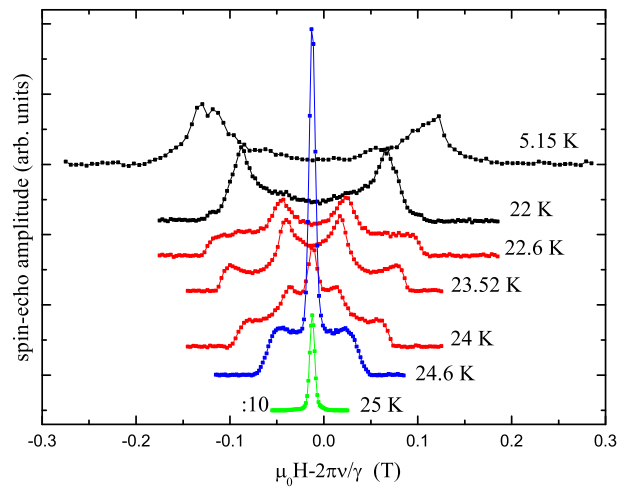}
\caption{(color online)
Temperature evolution of $^7$Li NMR spectra below the spin-reorientation transition, $\vect{H}\parallel \vect{a}$, $\nu = 127$~MHz, $2\pi\nu / \gamma = 7.68$~T. The spectra are in the paramagnetic (green), intermediate $T_{c2} < T < T_{c1}$ (red), and ordered phase $T < T_{c2}$ (black), respectively.
}
\label{fig:fig3}
\end{figure}

The magnetic field $\vect{H}_{eff}$ is defined by the magnetic environment.
If the effective field $H_{eff}$ varies harmonically in space and the wave length of this variation is incommensurate with the crystal lattice, we expect a broad NMR spectral line where its shape is dominated by two characteristic maxima at the edges. These maxima occur for the case $\vect{H}_{eff}\parallel \vect{H}$ ($H \gg H_{eff}$) and we designate this particular shape as a double-horn pattern.

\begin{figure}
\includegraphics[width=0.95\columnwidth,angle=0,clip]{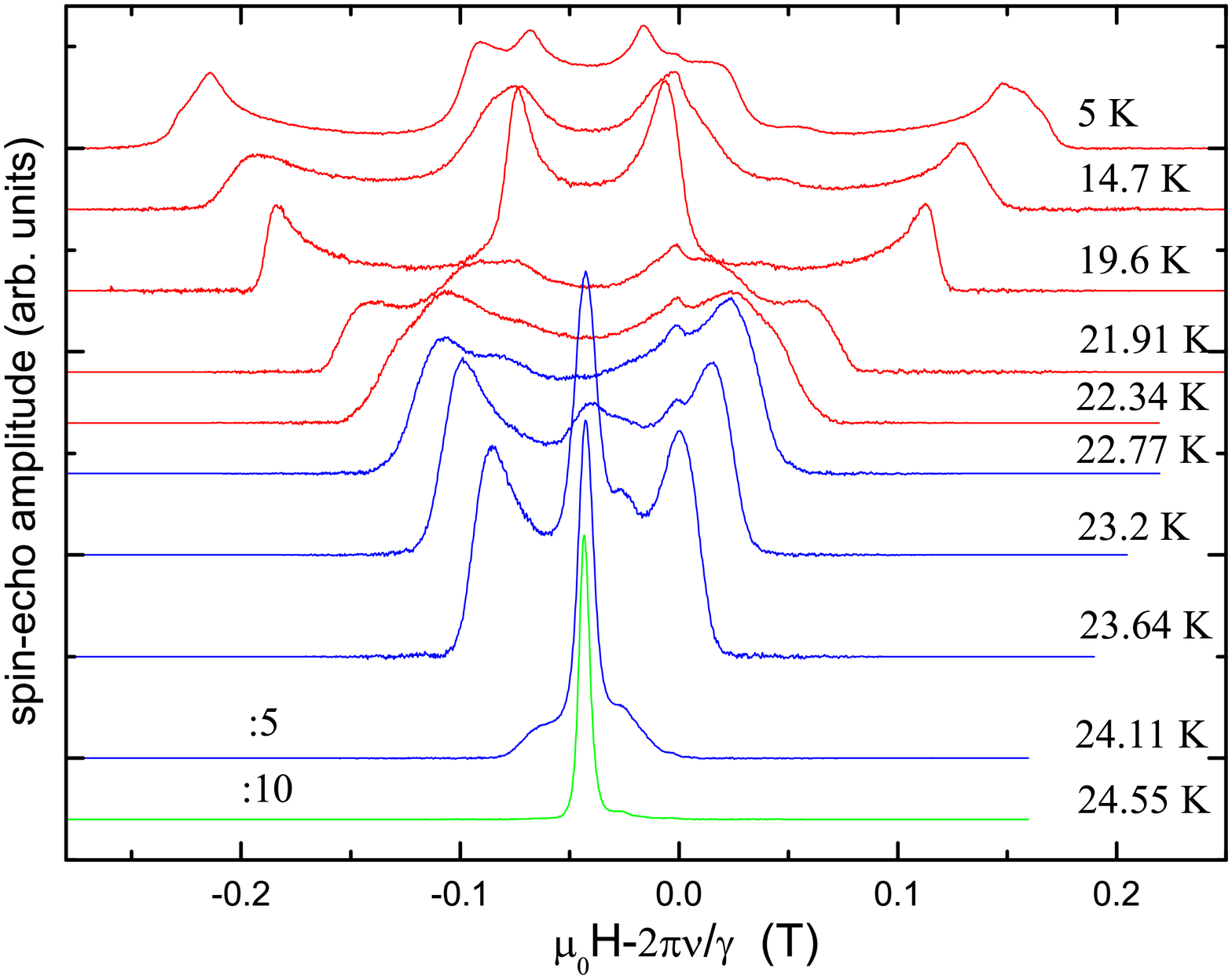}
\caption{(color online)
Temperature evolution of $^7$Li NMR spectra below the spin-reorientation transition, $\vect{H}\parallel \vect{a}$, $\nu = 279.178$~MHz, $2\pi\nu / \gamma = 16.87$~T. The spectra are in the paramagnetic (green), intermediate $T_{c2} < T < T_{c1}$ (blue), and ordered phase $T < T_{c2}$ (red), respectively.
}
\label{fig:fig4}
\end{figure}

There are eight chains of Li$^+$ ions along the $b$ direction in the magnetic structure of LiCu$_2$O$_2$ with the vector
$\vect{q} = (1/2, q, 0)$.~\cite{Svistov_2009} All other lithium chains can be obtained by translations. For each of the eight chains we expect a harmonically oscillating effective field generated by the magnetic surrounding.
Therefore, in the low-temperature magnetic phase the NMR spectra comprise a superposition of eight double-horn patterns, where some of them may coincide due to symmetry restrictions.
All the observed spectra are well described by a sum of not more than four double-horn spectra with nearly the same integral intensity. This fact shows that lithium chains, distanced by the lattice constant $a$, yield identical NMR spectra.

\begin{figure}
\includegraphics[width=0.95\columnwidth,angle=0,clip]{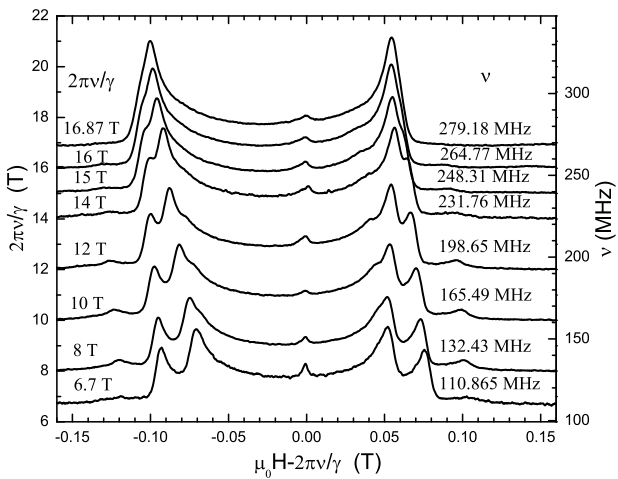}
\caption{
Field evolution of $^7$Li NMR spectra, $\vect{H}\parallel \vect{b}$, $T = 5$~K.
}
\label{fig:fig5}
\end{figure}

Figure~2 shows $^{7}$Li NMR field-swept spectra for $\vect{H}\parallel \vect{a}$. For the fields below
$\mu_0H_{c1}\approx 15$~T all four lithium chains demonstrate identical double-horn shaped lines. This means that the $a$-projections of effective fields on $^{7}$Li nuclei oscillate along the lithium chains with the same amplitudes. At higher fields the spectrum transforms into a sum of four double-horn shaped spectra. The value of $H_{c1}$ agrees with the field of an anomaly in the magnetization curves.~\cite{Svistov_2012} The step-like increase of the magnetic susceptibility at this field was previously associated with a spin-flop transition. The value of $H_{c1}$ is in satisfactory agreement with the value evaluated from ESR in an investigation of the spin-flop transition in LiCu$_2$O$_2$.~\cite{Svistov_2009} At fields nearby the phase transition the spectra exhibit hysteretic feature: The spectra recorded upon field increase differ from the spectra recorded upon field decrease (Fig.~2, $2\pi\nu / \gamma = 15$~T) which indicates that the phase transition observed at this field is of first order typical for spin-flop transitions.

\begin{figure}
\includegraphics[width=0.95\columnwidth,angle=0,clip]{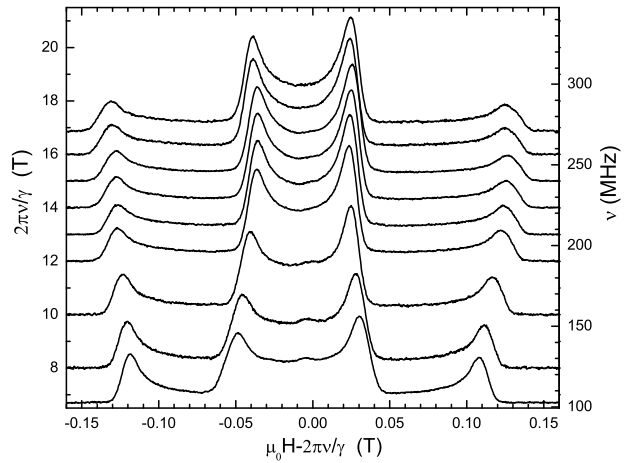}
\caption{
Field evolution of $^7$Li NMR spectra, $\vect{H}\parallel \vect{c}$, $T = 5$~K.
}
\label{fig:fig6}
\end{figure}

\begin{figure}
\includegraphics[width=0.95\columnwidth,angle=0,clip]{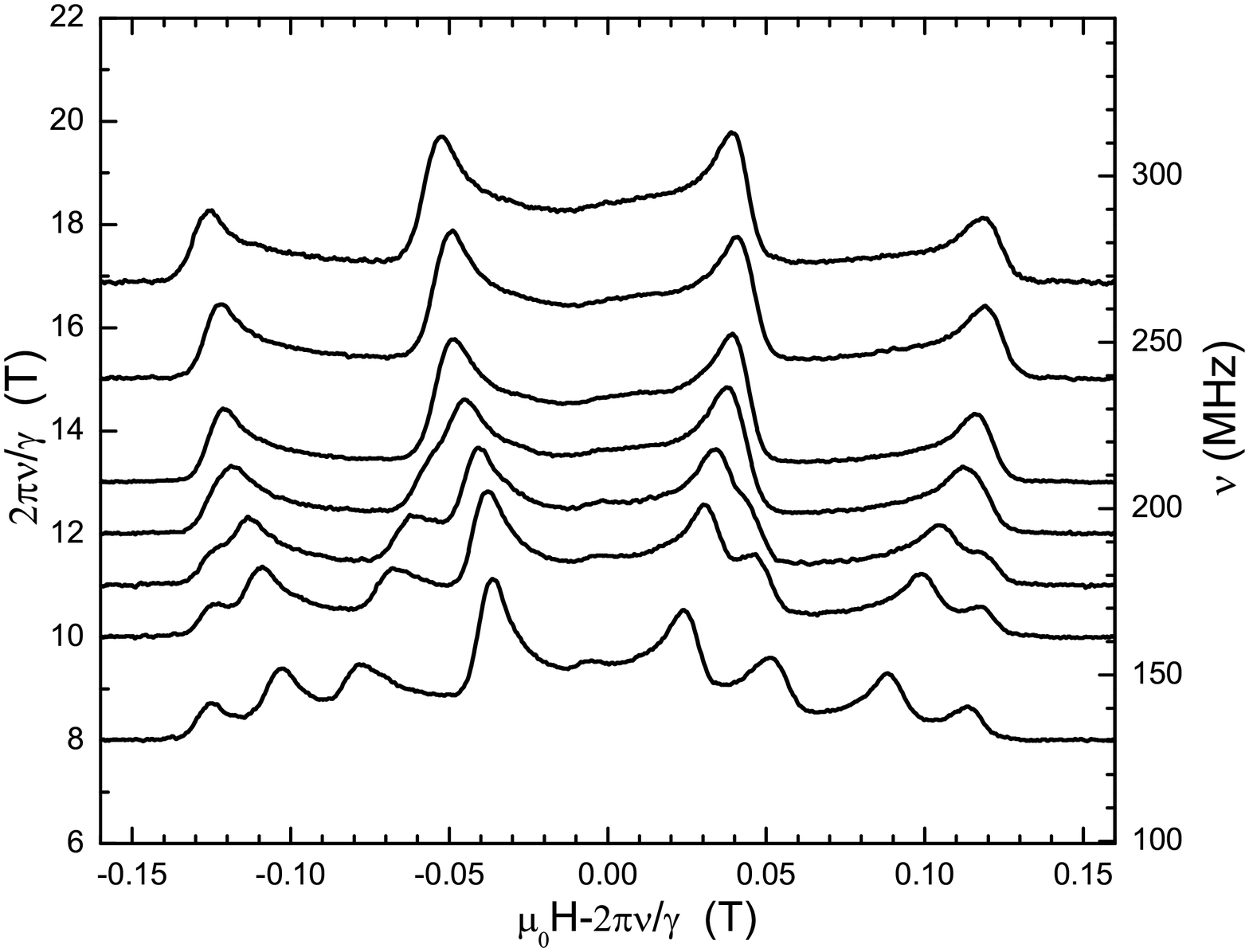}
\caption{
Field evolution of $^7$Li NMR spectra. $\vect{H}$ is located in $bc$-plane 15$^\circ$ off $c$ axis, $T = 5$~K.
}
\label{fig:fig7}
\end{figure}

Figures~3 and 4 show the temperature evolutions of NMR spectra measured around ($\mu_0H\approx 7.68$~T) and ($\mu_0H\approx 16.87$~T), i.e., below and above spin-flop field $H_{c1}$, respectively. The single line of the paramagnetic phase is significantly broadened in the magnetically ordered phase at $T_{c1} = (24.8 \pm 0.2)$ and $(24.3 \pm 0.2)$~K for 7.68 and 16.87~T, respectively. Within a range of 1~K below $T_{c1}$, the NMR spectrum can be considered as a superposition of a double-horn spectrum and a paramagnetic solitary line. These spectra are blue colored and observed between the temperatures $T_{c1}$ and $T_{c2}$ of the two step transition into the magnetically ordered phase. The intensity of the paramagnetic line rapidly decreases with decreasing temperature. At temperatures below $T_{c2}$, the double-horn spectrum transforms into two (four) double-horn spectra. These spectra are red colored in Figs.~3 and 4. We suggest that they are obtained within spin-flopped phase. The boundary between phases with blue and red colored spectra for the fields up to 9~T was presumably observed in Ref.~[\onlinecite{Svistov_2012}] by dielectric constant measurements.

\begin{figure}
\includegraphics[width=0.95\columnwidth,angle=0,clip]{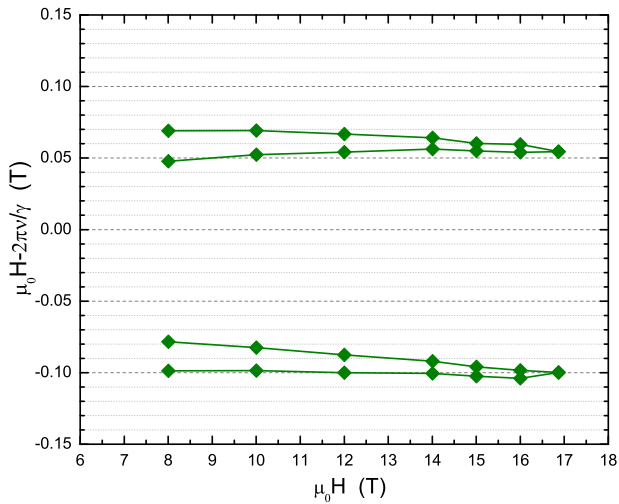}
\caption{
Peak positions of the spectra in Fig.~5.
}
\label{fig:fig8}
\end{figure}

\begin{figure}
\includegraphics[width=0.95\columnwidth,angle=0,clip]{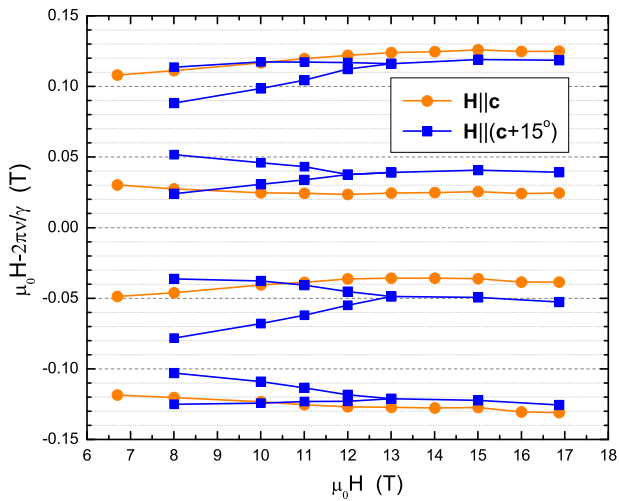}
\caption{
Peak positions of the spectra in Figs.~6 and 7.
}
\label{fig:fig9}
\end{figure}

Figures~5, 6, and 7 show the field evolutions of Li NMR spectra measured for the field directions $\vect{H}\parallel \vect{b}$, $\vect{H}\parallel \vect{c}$, and $\vect{H}\parallel (\vect{c}+15^\circ)$, respectively.
All spectra were measured at high enough fields, where the normal vector of the magnetically ordered spin plane appears to follow the field direction.~\cite{Svistov_2009} For $\vect{H}\parallel \vect{b}$ a continuous transformation from two double-horn shaped lines to one double-horn shaped line is observed upon field increase. For $\vect{H}\parallel \vect{c}$ the spectra do not change in the full range of applied magnetic fields as it is documented in Fig.~6.
An applied field direction with small deviation by 15$^\circ$ off the crystallographic $c$-axis yields a duplication of the number of double-horn pattern towards lower applied magnetic fields (see Fig.~7).
In other words, for this field orientation all four lithium ions show the double-horn spectrum with individual amplitude of oscillating projection $H_{eff}$. The field splitting decreases with increasing field and disappears at elevated fields higher than $(12.5 \pm 0.5)$~T.

Figures~8 and 9 show the field dependencies of the resonance field values of all edge singularities within the multi-horn spectral pattern. The fields where the number of edge singularities is halved are 17 and 12.5~T for $\vect{H}\parallel \vect{b}$ and $\vect{H}\parallel (\vect{c}+15^\circ)$, respectively, corresponding to the anomalies observed in the magnetization measurements.~\cite{Svistov_2012} The fields values of these anomalies were designated as $H_{c2}$. Such an anomaly was also observed in the magnetization measurements for $\vect{H}\parallel \vect{a}$ at $\mu_0H_{c2}\approx 20$~T, that is out of range of the present experiments.
It appears that we also observed this transition for $\vect{H}\parallel \vect{a}$ within the temperature scan at $\mu_0H\approx 16.87$~T (see Fig.~4). There, also the number of edge singularities halves for temperatures 15-20~K and the multi-horn shape of the spectra pattern in this temperature range strongly resembles to the shapes observed at lowest temperatures for $H > H_{c2}$.

\begin{figure}
\includegraphics[width=0.95\columnwidth,angle=0,clip]{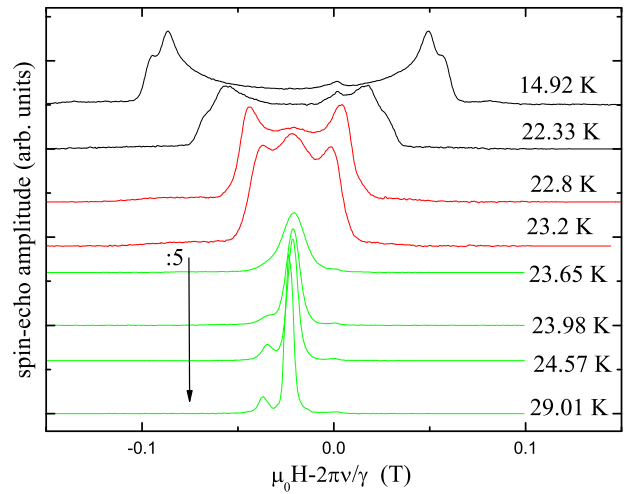}
\caption{(color online)
Temperature evolution of $^7$Li NMR spectra, $\vect{H}\parallel \vect{b}$, $\nu = 248.331$~MHz, and $2\pi\nu / \gamma = 15$~T.
}
\label{fig:fig10}
\end{figure}

\begin{figure}
\includegraphics[width=0.95\columnwidth,angle=0,clip]{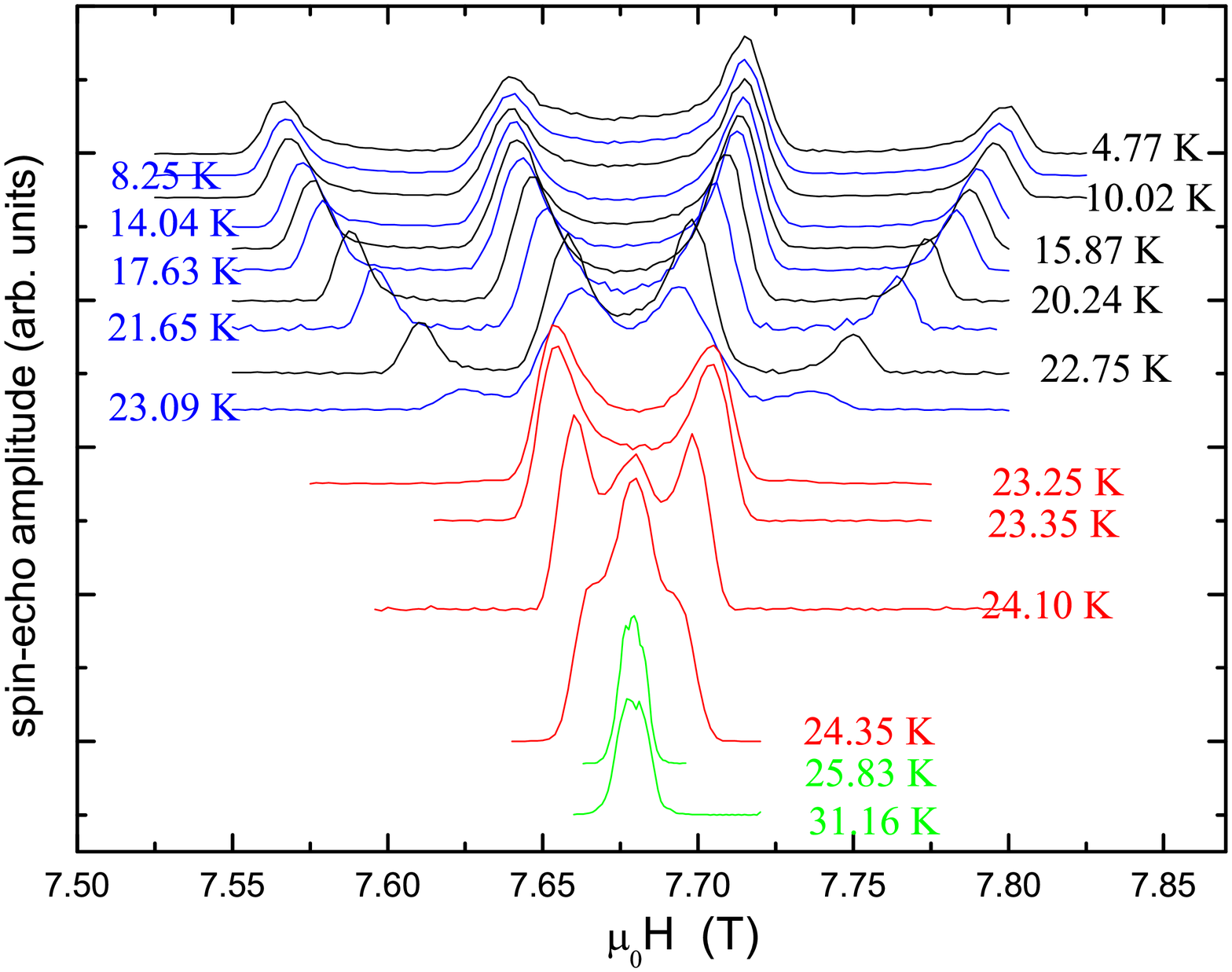}
\caption{(color online)
Temperature evolution of $^7$Li NMR spectra, $\vect{H}\parallel \vect{c}$, $\nu = 127$~MHz,
and $2\pi\nu / \gamma = 7.68$~T.
}
\label{fig:fig11}
\end{figure}

\begin{figure}
\includegraphics[width=0.95\columnwidth,angle=0,clip]{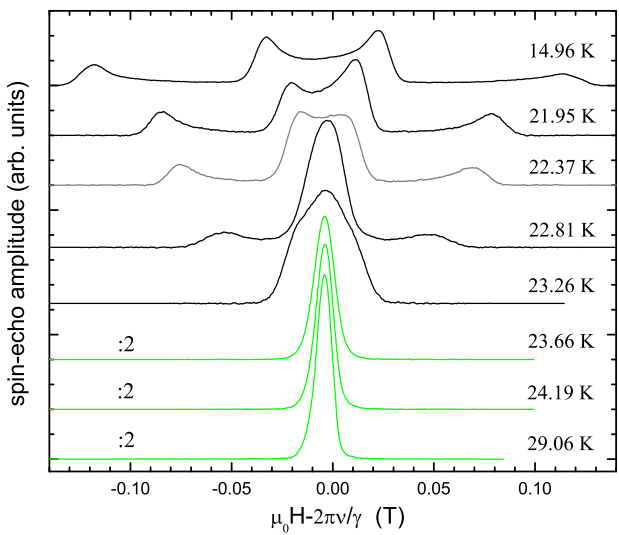}
\caption{(color online)
Temperature evolution of $^7$Li NMR spectra, $\vect{H}\parallel \vect{c}$, $\nu = 248.331$~MHz,
and $2\pi\nu / \gamma = 15$~T.
}
\label{fig:fig12}
\end{figure}

The temperature evolutions  of  the lithium spectra for ${\vect{H}\parallel\vect{b}}$ and $\vect{H}\parallel \vect{c}$ are given in Figs.~10-12. For fields lower than the critical field $H_{c2}$, the transition from the low-temperature magnetic phase to the paramagnetic phase occurs through an intermediate phase, with one solitary double-horn spectral line (see Fig.~11). This intermediate phase manifests itself within the temperature range between $T_{c1}$ and $T_{c2}$. For elevated fields higher than $H_{c2}$ (Fig.~12), the presence of the intermediate phase is not established.
In this case, the single line spectrum of the paramagnetic phase transforms immediately into the characteristic spectral pattern of the low-temperature phase, even within the temperature steps of our measurements. Any intermediate magnetically ordered phase is skipped.

\section{Theory. Exchange approximation} \label{Marchenko}

The crystal cell of LiCu$_2$O$_2$ contains four magnetic ions Cu$^{2+}$ ($S = 1/2$). At the magnetic transition a doubling of the period along the twofold $C_{2x}$ axis occurs, and as a result eight spiral spin chains appear. Their mutual orientation, amplitudes, phases and possible ellipticity are unknown.

As a first step to interpret the spin structure of LiCu$_2$O$_2$, it is useful to find out
the types of the structures with the wave vector $(1/2,q,0)$ occurring in the Dzyaloshinskii-Landau theory.
As it is usually done applying the Landau theory of second-order phase transitions, we assume the region of critical fluctuations to be small, which in our case fits with the experimental observations. In this case, the small value of the spin ($S = 1/2$) is not essential at all, since the quantum fluctuations near the transition are always small in comparison with the thermodynamic ones.

In this section we define a list of such structures -- candidates for description of the phase realized in LiCu$_2$O$_2$
at temperatures below $T_{c2}$, as well as within the intermediate phase ${(T_{c2}<T<T_{c1})}$.
It was found that the splitting of the transition is most likely due to small relativistic effects, and that even an additional transition is possible between the temperatures $T_ {c1}$ and $T_ {c2}$.

{\bf{The crystal symmetry group}} of LiCu$_2$O$_2$ -- $Pnma$  $(D_{2h}^{16})$ -- is defined by the three translations
${\tau_a:\,x\rightarrow x+1},$
${\tau_b:\,y\rightarrow y+1},$ ${\tau_c:\,z\rightarrow z+1}$, the inversion ${I:\,(x,y,z)\rightarrow(-x,-y,-z)}$ and
the two screw rotations \[C_{2x}:\, (x,y,z)\rightarrow(x+\frac{1}{2},-y+\frac{1}{2},-z+\frac{1}{2}),\]
\[C_{2y}:\,(x,y,z)\rightarrow(-x,y+\frac{1}{2},-z).\]

Two-dimensional complex representation corresponding to the wave vector $(1/2,q,0)$ (compare with the last case considered in Sec.~134 of the book Ref.~[\onlinecite{LL5}]) is implemented by the functions $\sin\pi xe^{\pm iqy},$ $\cos\pi xe^{\pm iqy}$ (the coordinates $x,y,z$ are measured in the unit cell parameters $a,b,c$, respectively). The functions that transform according to other possible representations associated with this wavevector differ by the factors which are insignificant for constructing the invariants even in the order parameter. For example, there is a representation with a pseudo-scalar
factor $\sin2\pi x\sin2\pi y\sin2\pi z$.

{\bf{The spin density}} arising at a second-order transition over the considered representation is
\begin{equation}\label{4f}\begin{split}
{\bf s}({\bf r})=\eta\{\sin\pi
x(\boldsymbol{\mu}e^{iqy}+\boldsymbol{\mu}^*e^{-iqy})+\\ +\cos\pi
x(\boldsymbol{\nu}e^{iqy}+\boldsymbol{\nu}^*e^{-iqy})\}f({\bf
r}),\end{split}\end{equation} where $\eta$ is the magnitude of the magnetic order parameter; $\boldsymbol{\mu}, \boldsymbol{\nu}$ are the complex vectors in the spin space, normalized by the condition ${\boldsymbol{\mu}\boldsymbol{\mu}^*+
\boldsymbol{\nu}\boldsymbol{\nu}^*=1}$; $f({\bf r})$ is a scalar function of coordinates which is invariant with
respect to the crystal symmetry group in paramagnetic phase.

To select the exchange effects, we suppose\cite{D} that the effect of crystalline transformations on the function ${\bf s}({\bf r})$ is reduced to a corresponding change of coordinates $ (x, y, z) $ at a fixed orientation of the spin space. In the Landau theory of second-order phase transitions, it is convenient to transfer
the laws of crystalline transformations from the coordinate functions to the coefficients
\begin{equation}\label{ICC}\begin{split}
I:&\, \boldsymbol{\mu}\rightarrow-\boldsymbol{\mu}^*,\,
\boldsymbol{\nu}\rightarrow\boldsymbol{\nu}^*;\\
C_{2x}:& \,\boldsymbol{\mu}\rightarrow
-e^{-iq/2}\boldsymbol{\nu}^*,\, \boldsymbol{\nu}\rightarrow
e^{-iq/2}\boldsymbol{\mu}^*;
\\ C_{2y}:&
\,\boldsymbol{\mu}\rightarrow-e^{iq/2}\boldsymbol{\mu},\,
\boldsymbol{\nu}\rightarrow e^{iq/2}\boldsymbol{\nu};\\
\tau_a:&\,\boldsymbol{\mu}\rightarrow-\boldsymbol{\mu},
\,\boldsymbol{\nu}\rightarrow-\boldsymbol{\nu};\\
\tau_b:&\,\boldsymbol{\mu}\rightarrow e^{iq}\boldsymbol{\mu},
\,\boldsymbol{\nu}\rightarrow e^{iq}\boldsymbol{\nu}.
\end{split}\end{equation}

For the considered representation, there is the Lifshitz exchange invariant
\begin{equation}\label{L}\begin{split}
\boldsymbol{\mu}\partial_x\boldsymbol{\nu}^*- \boldsymbol{\nu}^*\partial_x\boldsymbol{\mu}+
\boldsymbol{\mu}^*\partial_x\boldsymbol{\nu}- \boldsymbol{\nu}\partial_x\boldsymbol{\mu}^*,
\end{split}\end{equation}
leading to the instability of the phase transition.
Therefore observing a continuous transition over the representation with the wavevector $(1/2,q,0)$ in LiCu$_2$O$_2$ implies that the impact of the invariant (\ref{L}) in this antiferromagnet is small compared to the anisotropy effects, and we will not take it into account when considering the phases structure at ${T<T_{c2}}$.

In the exchange approximation, the Dzyaloshinskii-Landau expansion of free energy up to the fourth-order terms has the form   \begin{equation}\label{F}
F=\tau\eta^2 +\frac{\beta_0+{\cal{B}}}{2}\eta^4,\end{equation}
\begin{equation*}\begin{split}
{\cal{B}}=\beta_1(\boldsymbol{\mu}\boldsymbol{\nu}^*+
\boldsymbol{\mu}^*\boldsymbol{\nu})^2
+\beta_2(\boldsymbol{\mu}^*\boldsymbol{\mu}-
\boldsymbol{\nu}^*\boldsymbol{\nu})^2-\\
-\beta_3(\boldsymbol{\mu}\boldsymbol{\nu}^*-
\boldsymbol{\mu}^*\boldsymbol{\nu})^2+
\beta_4(\boldsymbol{\mu}^2+\boldsymbol{\nu}^2)
(\boldsymbol{\mu}^{*2}+\boldsymbol{\nu}^{*2})+\\
\beta_5(\boldsymbol{\mu}^2-\boldsymbol{\nu}^2)
(\boldsymbol{\mu}^{*2}-\boldsymbol{\nu}^{*2})+
4\beta_6(\boldsymbol{\mu}\boldsymbol{\nu})
(\boldsymbol{\mu}^*\boldsymbol{\nu}^*).
\end{split}\end{equation*}

Note that in the case of spin-1/2 the fourth order terms of the Landau expansion should be treated
as the result of thermodynamic averaging of the microscopic exchange Hamiltonian, which takes into
account simultaneous pair permutations of spins belonging to four or more atoms, for example,
a biquadratic term of the form $({\boldsymbol{\sigma}}_1 {\boldsymbol{\sigma}}_2) ({{\boldsymbol{\sigma}}_3 \boldsymbol{\sigma}}_4)$, where ${\boldsymbol{\sigma}}_i$ is the spin operator of the $i$th atom.
This peculiarity of the spin-1/2 case was established earlier in the study of the antiferromagnetic
phase of crystalline $^3$He in Ref.~\onlinecite{Roger_1983}.

The real form
$p=\boldsymbol{\mu}\boldsymbol{\nu}^*+\boldsymbol{\mu}^*\boldsymbol{\nu}$
is transformed as the $z$-component of the vector, and the real forms
$r=\boldsymbol{\mu\mu}^*-\boldsymbol{\nu\nu}^*,\, s=i\{\boldsymbol{\mu\nu}^*-\boldsymbol{\nu\mu}^*\}$
are transformed as components of deformation tensors $u_{xz}$ and
$u_{xy}$. For the rest spin convolutions we introduce the following notation:
$\zeta=\boldsymbol{\mu}^2+\boldsymbol{\nu}^2,\,
\omega=\boldsymbol{\mu}^2-\boldsymbol{\nu}^2,\,
\xi=\boldsymbol{\mu\nu}$.

The free energy $F$ does not change under the calibration transformation
\begin{equation}\label{imunu}\boldsymbol{\mu}\rightarrow e^{i\kappa}\boldsymbol{\mu}, \,
\boldsymbol{\nu}\rightarrow e^{i\kappa}\boldsymbol{\nu},
\end{equation}
which corresponds to the incommensurability of arising spin structure to the crystal spacing along the $y$ axis.
Using this invariance, we assume $\omega$ is real-valued.

The free energy $F$ is invariant under the replacement
$\boldsymbol{\mu}\leftrightarrows\boldsymbol{\nu}$,
which is connected with the solution transformation under the reflection ${\sigma_x=IC_{2x}}$.
Taking this into account, we will consider only the solutions with ${\boldsymbol{\mu}\neq0}$.

The free energy (\ref{F}) does not change under the transformation
\begin{equation}\label{inunu}\boldsymbol{\nu}\rightarrow i\boldsymbol{\nu}, \, \beta_1\leftrightarrows\beta_3, \, \beta_4\leftrightarrows\beta_5.
\end{equation}
Hence extrema that either remain unchanged or transform into each other under this transformation are possible.
Note that the Lifshitz invariant (\ref{L}) breaks this random symmetry.

Performing the transformation (\ref{inunu}) two times we obtain invariance $F$ under the replacement
$\boldsymbol{\mu}\rightarrow\boldsymbol{\mu}, \, \boldsymbol{\nu}\rightarrow-\boldsymbol{\nu}$.
This invariance is associated with the solution transformation under the crystal rotation $\tau_aC_{2y}$.

The extrema conditions of the free energy (\ref{F}) are
 \begin{equation}\label{mu}\begin{split}
-({\cal{B}}-\beta_2r)\boldsymbol{\mu}+(\beta_1p
-i\beta_3s)\boldsymbol{\nu}+\\+(\beta_4\zeta+
\beta_5\omega)\boldsymbol{\mu}^*+2\beta_6\xi\boldsymbol{\nu}^*=0,\end{split}\end{equation}
\begin{equation}\label{nu}\begin{split}
-({\cal{B}}+\beta_2r)\boldsymbol{\nu}+(\beta_1p
+i\beta_3s)\boldsymbol{\mu}+\\+(\beta_4\zeta-
\beta_5\omega)\boldsymbol{\nu}^*+2\beta_6\xi\boldsymbol{\mu}^*=0.
\end{split}\end{equation}

Performing scalar multiplication of the equation (\ref{mu}) by $\boldsymbol{\mu}^*$
and separating the imaginary part, we find
\begin{equation}\label{mumu**}
(\beta_1-\beta_3)ps+(\beta_4+
\beta_5)\zeta''\omega=0,\end{equation} where ${\zeta''=Im(\zeta)}.$

Multiplying the equation (\ref{nu}) by $\boldsymbol{\nu}^*$, for the imaginary part of the product we get
\begin{equation}\label{nunu*}
(\beta_1-\beta_3)ps+(\beta_4-\beta_5)\zeta''\omega=0.\end{equation}

From the equations (\ref{mumu**},\ref{nunu*}) it follows that
${ps=\zeta''\omega=0}.$ Hence in view of invariance of (\ref{inunu}), with ${s\rightarrow p\rightarrow-s},$ we can assume that, for example, ${p=0}.$
Thus, we have two cases
\begin{equation}\label{p}\begin{split}
1) \, p=\zeta''=0, \, 2) \, p=\omega=0.
\end{split}\end{equation}

At ${\omega=0}$ we can again use the calibration symmetry and choose $\zeta$ of real-valued. Thus, in all cases
it will be ${\zeta''=0}.$

Performing scalar multiplication of the equation (\ref{mu}) by $\boldsymbol{\mu}$ and separating the imaginary part
at ${p=\zeta''=0}$, we find
\begin{equation}\label{sxi}s\xi'=0.\end{equation}

Substraction of the equation (\ref{mu}) multiplied by $\boldsymbol{\nu},$ and the equation (\ref{nu}) multiplied
by $\boldsymbol{\mu},$ at ${p=0}$ gives
\begin{equation}\label{numu}
2(\beta_2-\beta_6)r\xi+i(-\beta_3+\beta_4)s\zeta=0.\end{equation}
Hence at ${\zeta''=0}$ we find ${r\xi'=0}.$ Combining this result with the conditions (\ref{p},\ref{sxi}) we obtain
the following possibilities
\begin{equation}\label{1-2}\begin{split}&1)\,p=\zeta''=r=s=0,\\
&2)\,p=\zeta''=\xi'=0.\end{split}\end{equation}

Let us introduce four real-valued vectors ${\bf a},{\bf b},{\bf c},{\bf d},$ such that ${\boldsymbol{\mu}={\bf a}+i{\bf b}},$ ${\boldsymbol{\nu}={\bf c}+i{\bf d}.}$ For all possible cases from the general conditions of reality of $\omega,$
$\zeta$ it follows, that $\boldsymbol{\mu}^2,$ $\boldsymbol{\nu}^2$ are real, which leads to the orthogonality condition
${{\bf ab}={\bf cd}=0}$.

According to the Appendix~A the magnetic structure should be planar and can therefore be written as:
\begin{equation}\label{par}\begin{split}
&\boldsymbol{\mu}=c_\alpha({\bf
l}c_\gamma+i{\bf k}s_\gamma),\\
&\boldsymbol{\nu}=s_\alpha\{c_\epsilon({\bf l}c_\varphi+{\bf
k}s_\varphi) -is_\epsilon({\bf l}s_\varphi-{\bf
k}c_\varphi)\},\end{split}\end{equation} where ${\bf l}, {\bf k}$
are mutually orthogonal unit vectors (we use short notation ${c_\alpha=\cos\alpha,}$ \, ${s_\alpha=\sin\alpha}$).

In such parametrization, we have ${p=s_{2\alpha}c_{\varphi}c_{-}},$
${r=c_{2\alpha},}$ ${s=-s_{2\alpha}s_{\varphi}s_{+}},$
${\xi'=s_{2\alpha}c_{\varphi}c_{+}},$ and
\begin{equation}\label{Bn}\begin{split}
{\cal{B}}=\beta_1s^2_{2\alpha}c^2_{\varphi}c^2_{-}+
\beta_2c^2_{2\alpha}+\beta_3s^2_{2\alpha}s^2_{\varphi}s^2_{+}\\
+\beta_4(c_{+}c_-{-}c_{2\alpha}s_{+}s_{-})^2\\
+\beta_5(c_{+}c_{-}c_{2\alpha}-s_{+}s_{-})^2\\
+\beta_6s^2_{2\alpha}(c^2_{\varphi}c^2_{+}+ s^2_{\varphi}s^2_{-})\end{split}\end{equation}
where${s_\pm=s_{\gamma\pm\epsilon},\,c_\pm=c_{\gamma\pm\epsilon}}.$
At the phase transition a spin structure corresponding to the minimum of ${\cal{B}}$ arises,
in this case the magnitude of the order parameter takes the maximum value ${\eta=\sqrt{-\tau/(\beta_0+{\cal{B}})}}.$
According to (\ref{1-2}), we get five scenarios for solving simultaneous equations (\ref{mu},\ref{nu})
\begin{equation}\label{AE}\begin{split}
A)&\, c_{2\alpha}=0, c_{\varphi}=s_+=0;\\
B)&\, c_{2\alpha}=0, s_{\varphi}=c_-=0;\\
C)&\, s_{2\alpha}=0;\\
D)&\, s_{2\alpha}\neq0, c_+=c_-=0;\\
E)&\, s_{2\alpha}\neq0, c_{\varphi}=0.\end{split}\end{equation}

In the framework of these scenarios solving is reduced to elementary minimization of the function (\ref{Bn})
over the remaining free angular parameters in each of them.~\cite{Redundant}
We find 8 solutions, whose form is independent of the values of $\beta_i:$
\begin{flalign*}
A_1:\boldsymbol{\mu}&=\frac{{\bf l}+i{\bf k}}{2},\,
\boldsymbol{\nu}={\pm}\frac{{\bf k}+i{\bf l}}{2},&{\cal{B}}=\beta_6;\\
A_2:\boldsymbol{\mu}&=\frac{\bf l}{\sqrt{2}},\,
\boldsymbol{\nu}=\frac{\bf k}{\sqrt{2}},&{\cal{B}}=\beta_4;\\
B_1:\boldsymbol{\mu}&=\frac{{\bf l}+i{\bf k}}{2},\,
\boldsymbol{\nu}={\pm}\frac{{\bf l}-i{\bf k}}{2},&{\cal{B}}=\beta_6;\\
B_2:\boldsymbol{\mu}&=\frac{{\bf l}}{\sqrt{2}},\,
\boldsymbol{\nu}=i\frac{\bf k}{\sqrt{2}},& {\cal{B}}=\beta_5;\\
C_1:\boldsymbol{\mu}&={\bf l},\,\boldsymbol{\nu}=0,&
{\cal{B}}=\beta_2+\beta_4+\beta_5;\\
C_2:\boldsymbol{\mu}&=\frac{{\bf l}+i{\bf k}}{\sqrt{2}},\,
\boldsymbol{\nu}=0,& {\cal{B}}=\beta_2;\\
D:\boldsymbol{\mu}&=\frac{\bf l}{\sqrt{2}},\,
\boldsymbol{\nu}={\pm}i\frac{\bf l}{\sqrt{2}},&
{\cal{B}}=\beta_3+\beta_5+\beta_6;\\
E:\,\boldsymbol{\mu}&=\frac{{\bf l}+i{\bf k}}{2},\,
\boldsymbol{\nu}={\pm}\frac{{{\bf k}-i\bf
l}}{2},& {\cal{B}}=\beta_3.
\end{flalign*}
\begin{equation}\label{ABCDE}%x
\end{equation}

The degeneracy of energy for the solutions $A_1$ and $B_1$ will be removed if we take into account the next terms of
the Dzyaloshinskii-Landau expansion. The solutions $A_1$ and $B_1$, as well as $A_2$ and $B_2$ are transformed into
each other under the transformation (\ref{inunu}). To the list of solutions we obviously need to add two more associated solutions:
\begin{flalign*}
D^*:\boldsymbol{\mu}&=\pm\boldsymbol{\nu}=\frac{\bf l}{\sqrt{2}}, &{\cal{B}}=\beta_1+\beta_4+\beta_6;\\
E^*:\,\boldsymbol{\mu}&=\pm\boldsymbol{\nu}=\frac{{\bf l}+i{\bf k}}{2},\,&{\cal{B}}=\beta_1.\end{flalign*}
\begin{equation}\label{D*E*}%x
\end{equation}

All the phases found can be presented on the phase diagram. Indeed, the value of ${\cal{B}}$ for each solution is obviously an absolute minimum of the function ${\cal{B}}$, if the parameters $\beta_i$ belonging to it are negative, and the rest of the parameters $\beta_i$ are positive.

Besides the magnetic structures (\ref{ABCDE},\ref{D*E*}) with fixed values of the phase parameters $\alpha,\gamma,\epsilon$, the symmetry allows the existence of magnetic structures, for which these parameters are functions of the coefficients of energy expansion $\beta_i$. These solutions may occur in a certain range of values of the parameters $\beta_i$ only in the scenario $E$, when $\varphi=\pi/2$, see equation (\ref{par}),
\begin{equation}\label{parE}
\boldsymbol{\mu}=c_\alpha({\bf
l}c_\gamma+i{\bf k}s_\gamma), \,
\boldsymbol{\nu}=s_\alpha(c_\epsilon{\bf
k} -is_\epsilon{\bf l}),\end{equation}
where the parameters $\alpha,\gamma,\epsilon$ are functions of $\beta_i$.
In this case rather complicated expressions occur, whose form is not necessary to determine the structure realized in
LiCu$_2$O$_2$ -- it is enough to have the formulas (\ref{parE}) with three free angles.

As an illustration, let us get the solution in the limit of large values of magnitude of the parameter $\beta_2$:
\begin{equation}\label{Eni}\begin{split}
c_+^2=\frac{\beta_3+\beta_5}{\beta_4+\beta_5}<1, c_-^2=\frac{\beta_6+\beta_5}{\beta_4+\beta_5}<1,\\ c_{2\alpha}=\frac{\beta_4+\beta_5}{\beta_2}c_+c_-s_+s_-,\\
{\cal{B}}\approx\beta_3s_+^2+\beta_4c_+^2c_-^2+\beta_5s_+^2s_-^2+\beta_6s_-^2.
\end{split}\end{equation}
There is also an associate of this solution, corresponding to the transformation (\ref{inunu}).

It should be noted that the Lifshitz exchange invariants (\ref{L}) occur in the phases $A_2, B_2, E, E^*$,
as well as in the set of phases described by (\ref{parE}).

In the presence of magnetic field there is an invariant
\begin{equation}\label{HH}\begin{split}
H_iH_k(\mu_i\mu_k^*+\mu_k\mu_i^*+\nu_i\nu_k^*+\nu_k\nu^*_i),
\end{split}\end{equation}
giving the anisotropy of the magnetic susceptibility tensor of exchange approximation.
For all the structures (\ref{ABCDE},\ref{D*E*}) this tensor has obvious axial symmetry. For the structures described by (\ref{parE}) the two main axes of the tensor belong to spin plane.

A spontaneous electric polarization of exchange nature as a quadratic effect in the order parameter (the effect predicted by Indenbom \cite{LL}) arises in the phases $D^*$ and $E^*$, where the quadratic form $p$ (transforming as $z$ component of vector) is not equal to zero.

Exchange striction corresponding to the symmetry breaking $D_{2h}$ arises in the phases $C_1,C_2$ (invariant $ru_{xz}$) and in the phases $D,E$ (invariant $su_{xy}$).

The  relativistic  effects  are  discussed  in  Appendix~B.

\section{Discussion}

\begin{figure*}
\includegraphics[width=2.05\columnwidth,angle=0,clip]{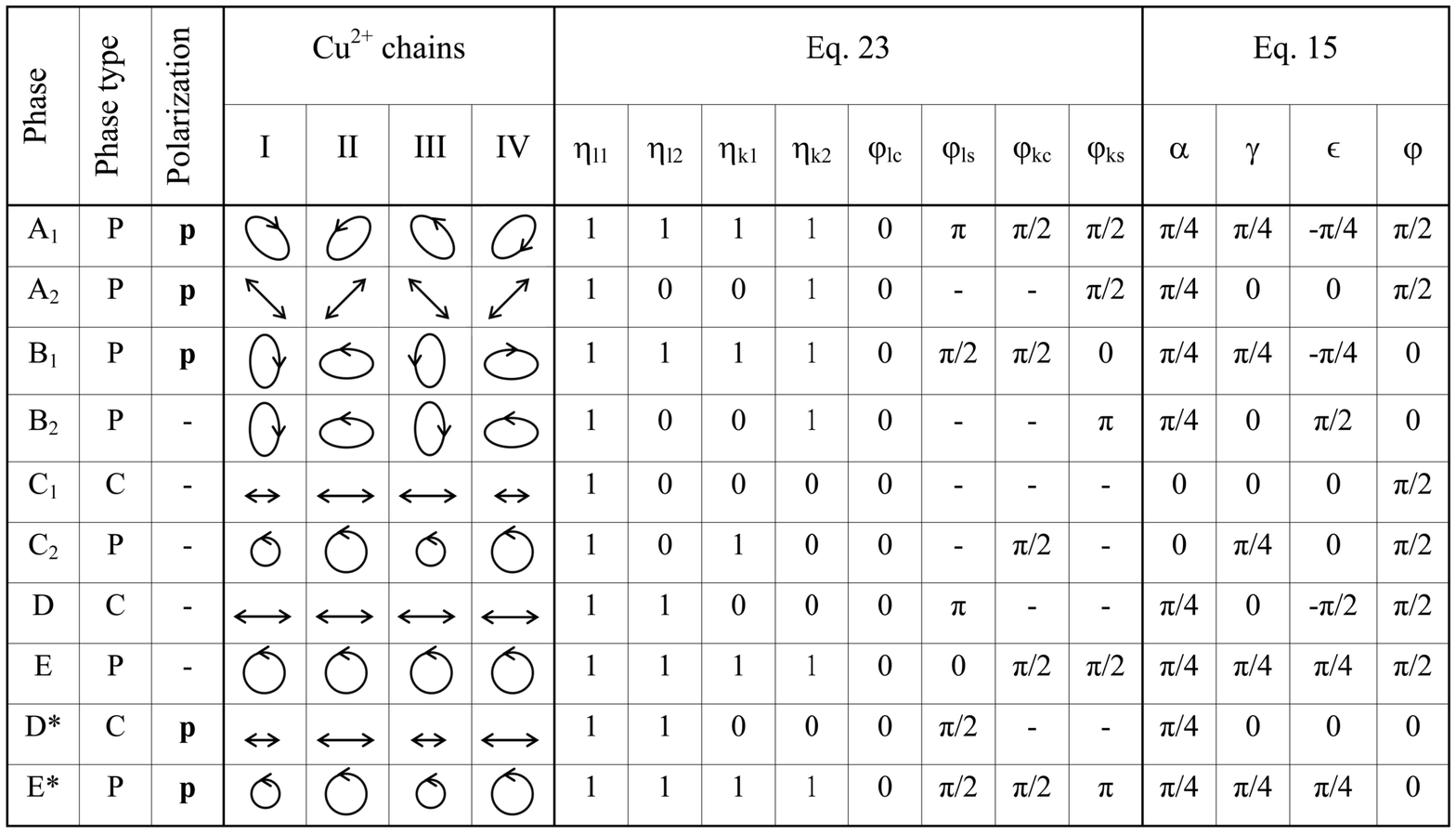}
\linebreak
\linebreak
Table~I. List of the phases Eqs.~(\ref{ABCDE}) and (\ref{D*E*}), their sketches and parameters in Eqs.~(\ref{par}) and (\ref{eqn:spiral}).
\end{figure*}

We discuss the observed NMR spectra using the magnetic phases obtained in Sec.~\ref{Marchenko}.
The theoretical analysis assumes that in the entire range of fields and temperatures the crystal structure of LiCu$_2$O$_2$ is described by the symmetry group $Pnma$, and the low-temperature magnetic structure is described by the wave vector $(1/2, q, 0)$. The proposed magnetic phases are obtained under the assumption that the magnetic structure is defined by the dominating exchange interactions, whereas its orientation with respect to the crystal axes is determined by the relativistic interactions with the crystal environment and the applied magnetic field. It was also demonstrated (see Appendix A) that the exchange interactions in LiCu$_2$O$_2$ can lead to planar or collinear magnetic structures only. Therefore, non-coplanar magnetic structures will be excluded from further consideration.

In the following we compare the experimental NMR spectra with the simulated ones.
We assume that the magnetic moments are localized at the Cu$^{2+}$ ions. This assumption accounts for the scalar function $f(\vect{r}) = \delta (\vect{r} - \vect{R}_i)$ in Eq.~(\ref{4f}), where $\vect{R}_i$ is the radius-vector of $i$th Cu$^{2+}$ ion. Using Eqs.~(\ref{4f}) and~(\ref{par}) the magnetic moment of Cu$^{2+}$ ion with coordinates $(x, y, z)$ for the coplanar magnetic structure with a wave vector $(1/2, q, 0)$ can be written as follows:

\begin{equation}
\begin{split}
\boldsymbol{M}(x,y,z)= \eta \Bigl\{
\vect{l}\Bigl[ \eta_{l1}\sin(\pi x)\cos(qy+\varphi_{lc}) + \\
+ \eta_{l2}\cos(\pi x)\sin(qy+\varphi_{ls})\Bigr] + \\
+ \vect{k}\Bigl[ \eta_{k1}\sin(\pi x)\cos(qy+\varphi_{kc}) + \\
+ \eta_{k2}\cos(\pi x)\sin(qy+\varphi_{ks})\Bigr]
\Bigr\}.
\end{split}
\label{eqn:spiral}
\end{equation}
Here, $\vect{l}$ and $\vect{k}$ are two mutually perpendicular unit vectors, defining the spin plane, and $\eta$ is the magnitude of the two component magnetic order parameter. The parameters $\eta_{l1}$, $\eta_{l2}$, $\eta_{k1}$ and $\eta_{k2}$ take the values 0 and 1. The angles $\varphi_{lc}$, $\varphi_{ls}$, $\varphi_{kc}$ and $\varphi_{ks}$ denote harmonic phase angles. According to Sec.~\ref{Marchenko} there are ten possible magnetic phases $A_1$, $A_2$, $B_1$, $B_2$, $C_1$, $C_2$, $D$, $E$, $D^*$, and $E^*$ with fixed phase parameters. An overview of all values of these parameters is given in Table~I.

The letters ``C'' and ``P'' in the table mark collinear and planar phases, respectively. The value of the magnetic moment for the collinear phases $C_1$, $D$, and $D^*$ oscillates harmonically along the chains. The phases $C_2$, $E$, and $E^*$ are circular. For them, the Cu$^{2+}$ magnetic moment with constant absolute value rotates by an angle defined by $q$.
Note, the absolute values of magnetic moments differ from chain to neighboring chains within the phases $C_2$ and $E^*$. The structures $A_1$, $B_1$, and $B_2$ can be considered as two embedded elliptical phases with large elliptical axes which are oriented perpendicular with respect to each other between neighboring axes. The planar structure $A_2$ consists of two embedded collinear structures in which the value of the magnetic moment varies along the chains. In this structure the magnetic moments of neighboring chains are oriented perpendicular with respect to each other. To visualize the structures, we included their sketches in the Table~I (the spin plane coincides with the easel plane, $\vect{l}$ is horizontal, $\vect{k}$ is vertical). The sketches show the ends of the magnetic moment vector for four chains I, II, III, and IV, the arrows indicate the moment rotation direction for the translation along the chain for one of magnetic domains. The magnetic phases which allow magnetically induced electrical polarization (multiferroicity) are marked with the letter ``p''.

Besides the structures given by Eqs.~(\ref{ABCDE}) and (\ref{D*E*}), there are the set of structures given by Eq.~(\ref{parE}). These structures are described by Eq.~(\ref{eqn:spiral}) with the following values of parameters: $\varphi_{lc} = \varphi_{ls} = 0$ and $\varphi_{kc} = \varphi_{ks} = \pi/2$. Other parameters are defined by:
$\eta_{l1} = 2 c_\alpha c_\gamma$, $\eta_{l2} = 2 s_\alpha s_\epsilon$,
$\eta_{k1} = 2 c_\alpha s_\gamma$, and $\eta_{k2} = 2 s_\alpha c_\epsilon$,
where $\alpha$, $\gamma$, and $\epsilon$ are arbitrary. The set of structures Eq.~(\ref{parE}) contains the phases $A_1$, $A_2$, $D$, and $E$ and adjoins to the phases $C_1$ and $C_2$.

Concluding the phase description, we note that all suggested magnetic phases are unusual. The values of the magnetic moments of all Cu$^{2+}$ ions are identical only for the $E$ phase, for other phases this value oscillates.

Space orientation of the spin structure is defined by the vectors $\vect{l}$ and $\vect{k}$. For further discussion we suppose that initially the vectors $\vect{l}$ and $\vect{k}$ are directed along $a$ and $b$ crystal axes, respectively. An arbitrary orientation of the spin plane can be obtained by successive rotation of the structure with
$\vect{l}\parallel \vect{a}$ and $\vect{k}\parallel \vect{b}$ by an angle $\Theta_a$ around $a$ axis, $\Theta_b$ around $b$ axis and $\Theta_c$ around $c$ axis. The number 1 or 2 or 3 in the brackets after the angle $\Theta$ specifies the order of the corresponding rotation where it matters. Thus the distribution of the magnetic moments is defined by the four parameters $\eta$, $\Theta_a$, $\Theta_b$, and $\Theta_c$ for the structures Eq.~(\ref{ABCDE},\ref{D*E*}) and by the seven parameters $\eta$, $\alpha$, $\gamma$, $\epsilon$, $\Theta_a$, $\Theta_b$, and $\Theta_c$ for the structures Eq.~(\ref{parE}). For large enough static fields ($H > H_{c1}$), the spin plane orientation is defined by the field direction, whereas in small fields the orientation is defined by relativistic interactions with the crystallographic environment.

In our spectra simulations we suppose that the effective field on lithium nuclei is defined by the dipolar fields and the contact Fermi fields. We took into account the dipolar fields from the neighboring moments in the sphere of radius 20~\AA{} and the hyperfine contact fields from the four nearest moments belonging to neighboring chains (Fig.~1). The constants defining the contact fields for three field orientations $\vect{H}\parallel \vect{a, b, c}$ were obtained as follows. First, the shift of $^7$Li NMR line at a certain temperature in the paramagnetic state was determined. Then, using the magnetization data from Ref.~[\onlinecite{Svistov_2012}], the magnetic moment of individual Cu$^{2+}$ ion at the same temperature and field was calculated. The difference between the computed dipole field at the lithium nuclei and the effective field observed in the experiment was ascribed to the contact field. Thus obtained contact field is in good agreement with the results of Ref.~[\onlinecite{Sadykov_2017}]. This value was ascribed to two pairs of the nearest magnetic moments. The first pair is located along $a$ axis, the second pair is located along $b$ axis (see Fig.~1). According to the first-principles calculations,~\cite{Sadykov_2017} the contact field from the first pair is expected to be much larger than the contribution from the second one. It is important that in the magnetically ordered phase the magnetic moments from the first pair are antiparallel; as a result, we can exclude the contact part of effective field in the magnetically ordered phase from further consideration.~\cite{Contact_field_note}

We have simulated the spectra for all magnetic structures from Table~I and orientations of the applied magnetic field
$\vect{H} \parallel \vect{a, b, c}$, and $(\vect{c}+15^\circ)$, respectively. For $\vect{H}\parallel \vect{a}$, the spectral shape does not change with applied magnetic fields up to $\approx 15$~T. For higher fields the susceptibility sharply increases, which was interpreted as a spin flop transition.~\cite{Svistov_2012} For all other field directions the shape changes with field monotonically. For $\vect{H} \parallel \vect{b}$ and $(\vect{c}+15^\circ)$ some pairs of maxima continuously approach each other and merge into one single spectral line at elevated applied fields. This disappearance of the line splitting is accomplished at a field value where anomalies of the magnetic susceptibility were already observed in Ref.~[\onlinecite{Svistov_2012}]. We tried to describe the field evolution of the spectra by changing the spin plane orientation of one individual magnetic phase. A special program was written for spectra simulation.~\cite{Program} It calculates NMR spectra for the suggested magnetic structures (\ref{ABCDE},\ref{D*E*},\ref{parE}) for given orientations of the static magnetic field with respect to the spin plane. The number of maxima in the spectrum was chosen as the main criterion in order to assess the matching between the observed NMR spectrum and the simulated one.
The calculations for all the structures Eqs.~(\ref{ABCDE}) and (\ref{D*E*}) performed with an angular step of 10$^\circ$ showed that the observed field evolution of the spectra for $\vect{H}\parallel \vect{a, b, c}$, and $(\vect{c}+15^\circ)$, respectively, can not be fitted assuming one single structure.

\begin{figure}
\includegraphics[width=0.95\columnwidth,angle=0,clip]{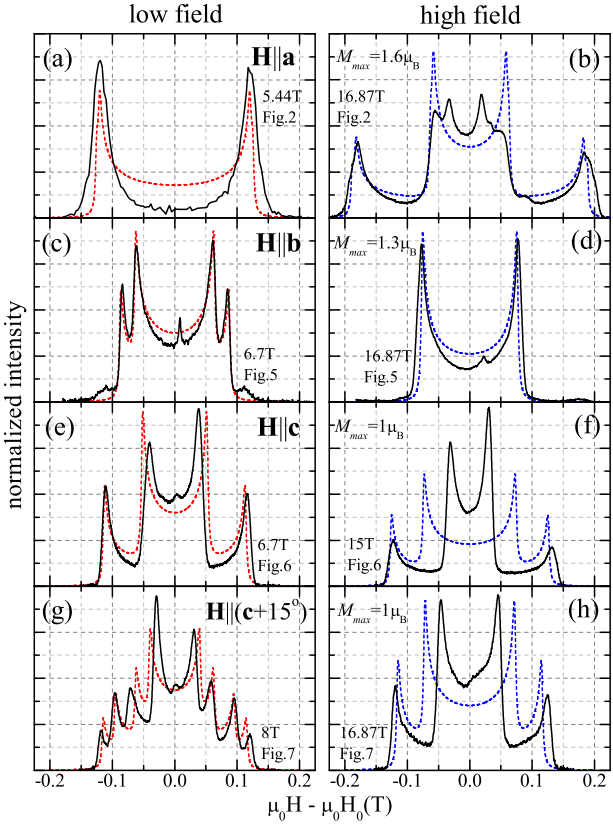}
\caption{(color online)
Dashed lines: Simulated NMR spectra, $M_{max}$ is noted in the figure, references to Fig.~14 for the structure sketches are given.
(a) $\vect{H}\parallel \vect{a}$, $H < H_{c1}$, $A_1$ structure, $\Theta_a = 90^\circ$, $\Theta_b = 0^\circ$, $\Theta_c = 0^\circ$, Fig.~14a;
(b) $\vect{H}\parallel \vect{a}$, $H > H_{c1}$, structure (\ref{parE}),
$\alpha = 72^\circ$, $\gamma = 60^\circ$, $\epsilon = 5^\circ$, $\Theta_a = 0^\circ$, $\Theta_b(2) = 90^\circ$, $\Theta_c(1) = 90^\circ$, Fig. ~14b;
(c) $\vect{H}\parallel \vect{b}$, $H < H_{c2}$, structure (\ref{parE}),
$\alpha = 20^\circ$, $\gamma = 45^\circ$, $\epsilon = -45^\circ$, $\Theta_a = 90^\circ$, $\Theta_b = 0^\circ$, $\Theta_c = 0^\circ$, Fig.~14c;
(d) $\vect{H}\parallel \vect{b}$, $H > H_{c2}$, structure (\ref{parE}),
$\alpha = 0^\circ$, $\gamma = 45^\circ$, $\epsilon = -45^\circ$, $\Theta_a = 90^\circ$, $\Theta_b = 0^\circ$, $\Theta_c = 0^\circ$, Fig.~14e;
(e) $\vect{H}\parallel \vect{c}$, $H < H_{c2}$, $A_1$ structure,
$\Theta_a = 0^\circ$, $\Theta_b = 0^\circ$, $\Theta_c = 0^\circ$, Fig.~14f;
(f) $\vect{H}\parallel \vect{c}$, $H > H_{c2}$, $A_1$ structure,
$\Theta_a = 0^\circ$, $\Theta_b = 0^\circ$, $\Theta_c = 45^\circ$, Fig.~14h;
(g) $\vect{H}\parallel (\vect{c}+15^\circ)$, $H < H_{c2}$, $A_1$ structure,
$\Theta_a = -15^\circ$, $\Theta_b = 0^\circ$, $\Theta_c = 0^\circ$, Fig.~14f;
(h) $\vect{H}\parallel (\vect{c}+15^\circ)$, $H > H_{c2}$, $A_1$ structure,
$\Theta_a(2) = -15^\circ$, $\Theta_b = 0^\circ$, $\Theta_c(1) = 45^\circ$, Fig.~14h.
Solid lines: Experimental NMR spectra, $T \approx 5$~K,  values of the applied magnetic field and its orientation are displayed additionally in each frame (a) to (h), see specified figures for details.
}
\label{fig:fig13}
\end{figure}

For $\vect{H}\parallel \vect{a}$ at the spin-flop field $H_{c1} \approx 15$~T the spin plane turns abruptly from the state with $\vect{n}\parallel \vect{b}$ ($\Theta_a = 90^\circ$, $\Theta_b = 0^\circ$, and $\Theta_c = 0^\circ$) to the state with $\vect{n}\parallel \vect{a}$ ($\Theta_a = 0^\circ$, $\Theta_b(2) = 90^\circ$, and $\Theta_c(1) = 90^\circ$). Figures~13(a) and 13(b) show simulated spectra for $H < H_{c1}$ and $H > H_{c1}$, respectively.
The best coincidence between experiment and simulation above the spin-flop transition is achieved if one assumes that a phase from the set Eq.~(\ref{parE}) is realized. Note here that the inflated value of the magnetic moment used for modeling at $\vect{H}\parallel \vect{a}$, in comparison with the expected value of 1~$\mu_B$, is most probably explained by the contact fields. This additional contribution to the local magnetic field at the $^{7}$Li nuclear site can be significant for the magnetic field applied in the spin plane.

\begin{figure}
\includegraphics[width=0.95\columnwidth,angle=0,clip]{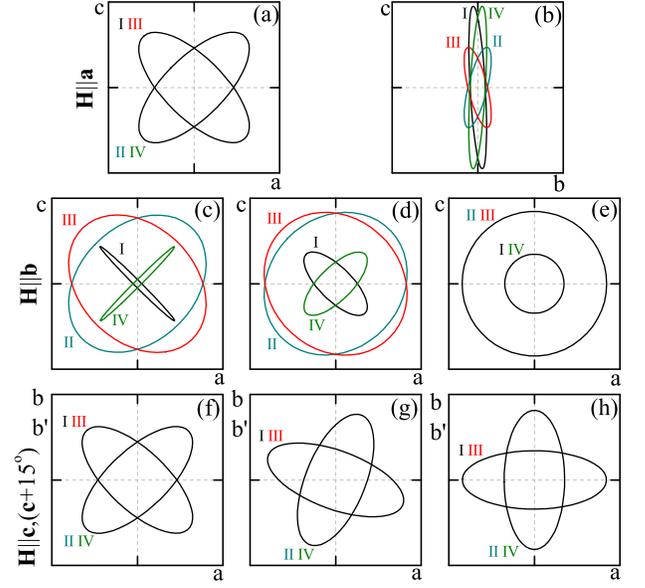}
\caption{(color online)
The sketches of the magnetic structures of four spirals I, II, III, IV. Shown are the projections of the magnetic moment vector endpoint on the corresponding spiral plane. The magnetic field orientation is shown at the left side. For all figures, $\vect{n} \parallel \vect{H}$ with the exception of (a). The $b'$ axis is $b$ axis rotated by an angle of -15$^\circ$
around the $a$ axis. Axes $b$ and $b'$ are used for $\vect{H}\parallel \vect{c}$ and $\vect{H}\parallel (\vect{c}+15^\circ)$, respectively. For the parameters of the structures and corresponding NMR spectra see Fig.13.
}
\label{fig:fig14}
\end{figure}

For $\vect{H}\parallel \vect{c}$ and $(\vect{c}+15^\circ)$, respectively, the spectra evolution can be modelled by rotation of the spin plane around the $c$ axis within the structure $A_1$. When $H > H_{c2}$ the spin plane rotation stops. For $\vect{H}\parallel \vect{c}$, the spectrum exhibits four maxima throughout the entire field range under investigation in this work. Figures~13(e) and 13(f) show simulated spectra for $H < H_{c2}$ and $H > H_{c2}$, respectively. For $\vect{H}\parallel (\vect{c}+15^\circ)$ in the low field range $H < H_{c2}$ the number of maxima doubles. When the field is increased to the value $\mu_0 H_{c2} \approx 12.5$~T the number of maxima becomes equal to four. Figures~13(g) and 13(h) show the simulated spectra for $H < H_{c2}$ and $H > H_{c2}$, respectively.

The orientation of the spin plane with respect to the crystallographic axes for different directions of the applied magnetic field for the phase $A_1$ can be explained by assuming the hierarchy of the anisotropy constants
$\lambda_2 > 0$, $\lambda_1 < 0$, $|\lambda_4| \ll |\lambda_2| < |\lambda_1|$.

For $\vect{H}\parallel \vect{b}$ we were unable to describe the field evolution of the spectra in the frame of our established structures which we introduce in Eqs.~(\ref{ABCDE}) and (\ref{D*E*}).
Figures~13(c) and 13(d) show a possible scenario in the frame of the phases Eq.~(\ref{parE}).
The low field phase exhibits a similar structure like the phase $A_1$.
As the field increases and gets closer to $H_{c2}$, the phase Eq.~(\ref{parE}) gradually transforms into the phase $C_1$. Both phases $A_1$ and $C_1$ are located on the boundary of the phases set described by Eq.~(\ref{parE}).

Fig.~14 shows the sketches of the structures used in the proposed scenario.

As a conclusion note, we cannot judge the uniqueness of the proposed scenario of the spectral field evolution within the magnetic structures set by Eq.~(\ref{parE}), since the number of free parameters defining these structures is too large.

As the temperature increases, the spectra shape does not change within a scale factor. This indicates that the magnetic structure is not changed until the temperatures approach the transition temperature, and the width of the spectra is determined by the order parameter $\eta$, see Eq.~(\ref{eqn:spiral}). The transition occurs through an intermediate phase observed in a narrow temperature range $T_{c2} < T < T_{c1}$. In this temperature range for all field orientations spectra with two characteristic maxima at the edges and a narrow central line ascribed to the paramagnetic state are observed. Relative intensities of these lines change with temperature. As the temperature increases, the intensity of the paramagnetic line increases, while the intensity of the double-horn pattern with the significant maxima at its edges decreases. A phase transition that takes place in two stages by passing through nearby transition temperatures is a characteristic fingerprint of a transition into a planar low-temperature phase in magnetic compounds
having easy-axis anisotropy. The observed spectra with such two characteristic maxima are well described by the collinear phase~$D$.

\section{Conclusions}

1. $^7$Li NMR spectra were measured in applied magnetic fields $\mu_0 H$ up to 17~T with the temperature range 5-30 K. Four different orientations of the field $\vect{H}\parallel \vect{a, b, c}$, and $(\vect{c}+15^\circ)$ with respect to the crystallographic axes where employed. The field dependent evolution of the spectra was studied in full detail. Anomalies in bulk measurements of the magnetization curves at fields $H_{c1}$ and $H_{c2}$, recently reported in the literature, were found to correspond to the fields values where the shape of our NMR spectra changes.
At $H_{c1}$,  a first-order phase transition is observed for $\vect{H}\parallel \vect{a}$,
which is accompanied by the increase of magnetic susceptibility and most likely indicates a spin-reorientation transition. The observation of the spin-reorientation transition for $\vect{H}\parallel \vect{a}$ supports the model of a planar two-component magnetic structure with the magnetic susceptibility in the field perpendicular to the spin plane larger than that in the field directed within the spin plane. This result is also in agreement with the biaxial character of the magnetic anisotropy suggested in Refs.~[\onlinecite{Svistov_2009, Svistov_2012}].
For other directions of the magnetic field at $H < H_{c2}$, the shape of the spectra monotonically evolves to a state with less number of maxima upon field increase. At the point where the evolution stops the magnetic susceptibility decreases. Such monotonous field evolution could be explained by the rotation of the spin plane or by monotonous change of magnetic structure. The modeling of our experimental NMR spectra shows that  the second scenario is realized.

2. We analyzed the underlying magnetic structures in the frame of Dzyaloshinskii-Landau theory of magnetic phase transitions. A set of possible planar and collinear structures for low temperatures and close to the transition temperature was obtained. Most of these structures have an unusual configuration. They are characterized by a two-component magnetic order parameter and their magnetic moments vary harmonically not only in direction, but also in size. A theoretical analysis made within the exchange approximation, revealed a set of magnetic structures given by Eqs.~(\ref{ABCDE}) and (\ref{D*E*}), which are determined by fixed spin configurations within each individual chain and fixed inter-chain phase relations. Besides these solutions a possibility that the minimum of exchange energy is achieved for a set of magnetic structures described by Eq.~(\ref{parE}) was found. For such a case, the configuration of the magnetic structure depends on the values of the coefficients of the energy expansion written in Eq.~(\ref{F}), $\beta_i$. For these structures, monotonous temperature and field evolutions of the spectra are expected and degeneracy removing occurs due to weak interactions. For such structures, the application of a magnetic field can lead not only to a rotation of the spin plane, but also yields a change into a structure out of the set which we present in Eq.~(\ref{parE}).

3. Our simulation of the experimental NMR spectra was performed ab initio from the structures
in Eqs.~(\ref{ABCDE}), (\ref{D*E*}), and (\ref{parE}). It has been shown
that the experimentally observed spectra for $\vect{H}\parallel \vect{a, b, c}$, and $(\vect{c}+15^\circ)$,
respectively, and for temperatures $T \ll T_N$, and applied fields $H \ll H_{sat}$ can only be described in the frame of the structures given in Eq.~(\ref{parE}). This result allowed us to propose a possible scenario of the magnetic structure evolution which should easily be confirmed by elastic neutron scattering experiments.

\acknowledgments

We gratefully acknowledge support by the German Research Society (DFG) via TRR80 (Augsburg, Munich).
The work was supported by the Program of the Presidium of Russian Academy of Sciences and by Russian Foundation for Basic Research (Grant No. 16-02-00688). A. A. Gippius acknowledge support from the RFBR BRICS Grant No. 17-52-80036.

\section{Appendix A}
Let us write the real and imaginary parts of the equation obtained by
scalar multiplication of the equations (\ref{mu},\ref{nu}) by ${{\bf
e}=[\bf{ab}]}$, ${{\bf g}=[\bf{cd}]}$:
\begin{equation}\label{ed}\begin{split}
(-{\cal{B}}+\beta_2r+\beta_4\zeta+
\beta_5\omega){\bf ag}=0,\\
(-{\cal{B}}+\beta_2r-\beta_4\zeta-
\beta_5\omega){\bf bg}=0,\\
(\beta_1p+2\beta_6\xi'){\bf ag}
+(2\beta_6\xi''-\beta_3s){\bf bg}=0,\\
(2\beta_6\xi''+\beta_3s){\bf ag}+(\beta_1p-2\beta_6\xi'){\bf bg}=0,\\
(-{\cal{B}}-\beta_2r+\beta_4\zeta-\beta_5\omega){\bf ce}=0,\\
(-{\cal{B}}-\beta_2r-\beta_4\zeta+\beta_5\omega){\bf de}=0,\\
(\beta_1p+2\beta_6\xi'){\bf ce}+(2\beta_6\xi''+\beta_3s){\bf de}=0,\\
(2\beta_6\xi''-\beta_3s){\bf ce}+(\beta_1p-2\beta_6\xi'){\bf de}=0.\\
\end{split}\end{equation}
These simultaneous equations lead to four types of solutions
(accurate within the above mentioned transformations
${\boldsymbol{\mu}\leftrightarrows\boldsymbol{\nu}},$
${\boldsymbol{\mu}\rightarrow\boldsymbol{i\mu}}$):

I. If ${{\bf ag}={\bf bg}=0},$ then the vectors ${\bf a},{\bf b},{\bf
c},{\bf d}$ are coplanar and therefor ${{\bf ce}={\bf
de}=0}.$ These structures were considered above.

II. If ${{\bf ag}={\bf ce}=0},$  ${{\bf bg},{\bf de}\neq0},$
i.e. when simultaneously three vectors ${\bf a},{\bf c},{\bf d}$ are coplanar
and three vectors ${\bf a},{\bf b},{\bf c}$ are coplanar,
this is possible only at ${\bf a}||{\bf c}$. Then ${p=s=\xi=0},$ ${\beta_2r=-\beta_5\omega}$, and
${{\cal{B}}=-\beta_4\zeta}.$

III. If ${{\bf ce}=0},$  ${{\bf ag},{\bf bg},{\bf de}\neq0},$ then
${\beta_1^2p^2+\beta_3^2s^2=4\beta_6^2\xi\xi^*}$ and
${{\cal{B}}=\beta_2r=-\beta_4\zeta=\beta_5\omega}.$

IV. If ${{\bf ag},{\bf bg},{\bf ce},{\bf de}\neq0},$ then
${r=\zeta=\omega={\cal{B}}=0},$
${\beta_1^2p^2+\beta_3^2s^2=4\beta_6^2\xi\xi^*}.$

Thus, the possibility of existence analysis for non-coplanar structures
is reduced to the analysis of 2 * 3 = 6 cases. Here the same solution can be obtained repeatedly, but no one solution will be missed.

Let us introduce two orthonormal bases in the spin space
${\bf l},{\bf k},{\bf n}$ and $\tilde{\bf l},\tilde{\bf k},\tilde{\bf n},$
and the parameters $\alpha\, \gamma, \epsilon,$ such that
\[{\bf a}={\bf l}c_{\alpha}c_\gamma,\,{\bf b}={\bf k}c_{\alpha}s_\gamma,\]
\[{\bf c}=\tilde{\bf l}s_{\alpha}c_\epsilon,\,
{\bf d}=\tilde{\bf k}s{\alpha}s_\epsilon.\]

Let us introduce Euler angles $\theta, \varphi, \psi,$
to specify the mutual orientation of the spin bases:
\begin{equation*}\begin{split}
&\tilde{\bf l}={\bf l}(c_\psi c_\varphi-
s_\psi c_\theta s_\varphi)+ {\bf k}(c_\psi s_\varphi+s_\psi
c_\theta c_\varphi)+ {\bf n}s_\psi s_\theta,\\
&\tilde{\bf k}=-{\bf l}(s_\psi c_\varphi+c_\psi c_\theta s_\varphi)
 +{\bf k}(c_\psi c_\theta c_\varphi-s_\psi s_\varphi)+
 {\bf n}c_\psi s_\theta,\\
&\tilde{\bf n}={\bf l}s_\theta s_\varphi-{\bf k}s_\theta c_\varphi+
{\bf n}c_\theta.
\end{split}\end{equation*}

II. From the condition ${\bf a}||{\bf c}$ it follows, that
\[\boldsymbol{\mu}=c_\alpha({\bf l}c_\gamma+i{\bf k}s_\gamma),\,
\boldsymbol{\nu}=s_\alpha\{c_\epsilon{\bf l}
+is_\epsilon({\bf k}c_\theta+{\bf n}s_\theta)\}.\]
In this case, the condition ${s=0}$ is performed automatically.
The condition ${p=\xi=0}$ for the considered non-coplanar structures (${c_\alpha s_\alpha\neq0}$)
is reduced to the relation ${c_\gamma c_\epsilon=s_\gamma s_\epsilon c_\theta=0}.$
Hence for non-coplanar structures we have either ${c_\gamma=c_\theta=0},$ or
${c_\epsilon=c_\theta=0}.$ In the first case
\[\boldsymbol{\mu}=ic_\alpha{\bf k},\,
\boldsymbol{\nu}=s_\alpha(c_\epsilon{\bf l} +is_\epsilon{\bf
n}),\] \[{\cal{B}}=\beta_2c^2_{2\alpha}+\beta_4(s^2_{\alpha}
c_{2\epsilon}-c^2_\alpha)^2+
\beta_5(s^2_{\alpha}c_{2\epsilon}+c^2_\alpha)^2.\]
It is easy to verify that there are no non-coplanar solutions here.
The same result is obtained in the case of ${c_\epsilon=c_\theta=0}.$

III. From the condition ${{\bf ce}=0}$ for non-coplanar structures it follows that ${s_{\psi}c_{\epsilon}=0}.$

IIIA. If ${s_{\psi}=0},$ then
\[\boldsymbol{\mu}=c_\alpha({\bf l}c_\gamma+i{\bf k}s_\gamma),\]
\[\boldsymbol{\nu}=s_\alpha\{c_\epsilon({\bf l}c_\varphi+ {\bf
k}s_\varphi)+is_\epsilon(-{\bf l}c_{\theta}s_\varphi+{\bf
k}c_{\theta}c_\varphi+{\bf n}s_\theta)\}.\] From the general for 1a,b
condition ${p=0}$ we found, that either ${c_\varphi=0},$ and then ${\bf a}||{\bf c},$ i.e.
we return to the already considered type of solution II, or
\begin{equation}\label{c-} c_\gamma c_\epsilon=-s_\gamma
s_\epsilon c_\theta.\end{equation}

In the case 1a) we have  ${r=0},$ whence it follows, that $c_{2\alpha}=0,$ i.e.
\[\boldsymbol{\mu}=\frac{{\bf l}c_\gamma+i{\bf k}s_\gamma}{\sqrt{2}},\]
\[\boldsymbol{\nu}=\frac{c_\epsilon({\bf l}c_\varphi+ {\bf
k}s_\varphi)+is_\epsilon(-{\bf l}c_{\theta}s_\varphi+{\bf
k}c_{\theta}c_\varphi+{\bf n}s_\theta)}{\sqrt{2}},\] and, according to III, ${\zeta=\omega=0}.$
Hence we found
${c_{2\gamma}=c_{2\epsilon}=0},$ i.e., according to (\ref{c-}),
${c_\theta=\cot\gamma \cot\epsilon=\pm 1},$ i.e. we return to coplanar structure again.

In the case 1b) from the condition ${\xi'=0}$ we find
\begin{equation}\label{c+} c_\gamma c_\epsilon=s_\gamma
s_\epsilon c_\theta.\end{equation}
For non-coplanar structure ${s_{\epsilon}\neq0},$
and from the equations (\ref{c-},\ref{c+}) it follows,
that ${c_{\gamma}c_\epsilon=s_{\gamma}c_\theta=0}.$ Hence
there are three options of solution:

$1. {c_{\gamma}=c_\theta=0:\,\boldsymbol{\mu}=ic_\alpha{\bf k},}$
\[\boldsymbol{\nu}=s_\alpha\{c_\epsilon({\bf l}c_\varphi+ {\bf
k}s_\varphi)-is_\epsilon{\bf n}\},\]
\[{\cal{B}}=\beta_2c^2_{2\alpha}+
(\beta_3+\beta_6)s^2_{2\alpha}c^2_{\epsilon}s^2_{\varphi}+\]
\[\beta_4(s^2_{\alpha}c_{2\epsilon}-c^2_\alpha)+
\beta_5(s^2_{\alpha}c_{2\epsilon}+c^2_\alpha);\]

$2. {c_\epsilon=s_{\gamma}=0}:$ ${\boldsymbol{\mu}=c_\alpha{\bf l}}$,
\[\boldsymbol{\nu}=is_\alpha\{c_{\theta}(-{\bf l}s_\varphi+{\bf
k}c_\varphi)+{\bf n}s_\theta\},\]
\[{\cal{B}}=(\beta_2+\beta_4)c^2_{2\alpha}+
(\beta_3+\beta_6)s^2_{2\alpha}c^2_{\theta}s^2_{\varphi}+\beta_5;\]

$3. {c_\epsilon=c_\theta=0:\,\boldsymbol{\nu}=is_\alpha{\bf
k},\,\boldsymbol{\mu}=c_\alpha({\bf l}c_\gamma+i{\bf
k}s_\gamma),}$
\[{\cal{B}}=\beta_2c^2_{2\alpha}+\beta_4(c^2_{\alpha}c_{2\gamma}+s^2_\alpha)^2
+\beta_5(c^2_{\alpha}c_{2\gamma}-s^2_\alpha)^2.\]
It is easy to verify that in these three options only coplanar structures appear.

IIIB. If ${c_{\epsilon}=0},$ then the vector ${{\bf c}=0}.$ Let us direct
$\tilde{\bf n}$ along the vector ${\bf d},$ then
\[\boldsymbol{\mu}=c_\alpha(c_\gamma{\bf l}+is_\gamma{\bf k})\]
\[\boldsymbol{\nu}=is_{\alpha}({\bf l}s_\theta s_\varphi-
{\bf k}s_\theta c_\varphi+ {\bf n}c_\theta),\]
\[{\cal{B}}=(\beta_1+\beta_3)c^2_{\gamma}s^2_{2\alpha}s^2_{\theta}c^2_\varphi+
(\beta_3+\beta_6)c^2_{\gamma}s^2_{2\alpha}s^2_{\theta}s^2_\varphi+\]
\[\beta_2c^2_{2\alpha}+
\beta_4(c^2_{\alpha}c_{2\gamma}-s^2_\alpha)^2+
\beta_5(c^2_{\alpha}c_{2\gamma}+s^2_\alpha)^2.\]
Here also only coplanar structures appear.

IV. From the condition ${r=\zeta=\omega=p=0}$ it follows, that
${\alpha=\gamma=\epsilon=\pi/4};\,\psi+\varphi=\pi/2.$ Thus,
${2\boldsymbol{\mu}={\bf l}+i{\bf k}}$ and
\[2\boldsymbol{\nu}=
{\bf l}s_{\varphi}c_{\varphi}(1-c_\theta)+ {\bf k}(s^2_\varphi+c_{\theta}c^2_\varphi)+ {\bf n}c_{\varphi}s_\theta,\]
\[+i\{-{\bf l}(c^2_\varphi+c_{\theta}s^2_\varphi)
 +{\bf k}s_{\varphi}c_{\varphi}(c_\theta-1)+
 {\bf n}s_{\varphi}s_\theta,\}\]
\[{\cal{B}}=\frac{\beta_3}{4}(1+c_\theta)^2+
\frac{\beta_6}{4}(1-c_\theta)^2.\] Minimizing the function $B$, we found
\[\,c_\theta=\frac{\beta_6-\beta_3}{\beta_6+\beta_3},\,
{\cal{B}}=\frac{\beta_3\beta_6}{\beta_3+\beta_6.}\]
We see, that the condition ${{\cal{B}}=0}$, as it should be
for the solutions type IV, is only performed when $\beta_3$ or $\beta_6$ equals to zero.

Thus, there are no extrema of the function ${\cal{B}},$ corresponding to
non-coplanar structures.

\section{Appendix B}

{\bf{Relativistic effects}}. To obtain relativistic invariants the action of rotational elements, in contrast to the exchange symmetry transformations (\ref {ICC}), has to be extended to the spin indices
\begin{equation}\label{CC}\begin{split}
C_{2x}:\, &\mu_x\rightarrow-e^{-iq/2}\nu_x^*,\, \mu_{y,z}\rightarrow e^{-iq/2}\nu_{y,z}^*,\\
&\nu_x\rightarrow e^{-iq/2}\mu_x^*, \nu_{y,z}\rightarrow-e^{-iq/2}\mu_{y,z}^*;\\ C_{2y}:\, &\mu_y\rightarrow-e^{iq/2}\mu_y,\, \mu_{x,z}\rightarrow e^{iq/2}\mu_{x,z},\\
&\nu_y\rightarrow e^{iq/2}\nu_y,\, \nu_{x,z}\rightarrow-e^{iq/2}\nu_{x,z}.
\end{split}\end{equation}

Anisotropy energy in general case has the form
\begin{equation}\label{Fan}\begin{split}
\lambda_1(\mu_x\mu^*_x+\nu_x\nu^*_x)+ \lambda_2(\mu_y\mu^*_y+\nu_y\nu^*_y)+\\
+\lambda_3(\mu_x\mu^*_z+\mu^*_x\mu_z-\nu^*_x\nu_z-\nu_x\nu^*_z)+\\
+i\lambda_4(\mu_x\nu^*_y-\mu^*_x\nu_y+\nu^*_x\mu_y-\nu_x\mu^*_y),
\end{split}\end{equation}
here the common factor $\eta^2$ has been omitted.

Let us give the expressions for the anisotropy energy of the phases:
\begin{equation*}\begin{split}
&A_1: \, -\frac{\lambda_1}{2}n^2_x-\frac{\lambda_2}{2}n^2_y\pm\lambda_4(l_xl_y-k_xk_y);\\
&A_2: \, -\frac{\lambda_1}{2}n^2_x-\frac{\lambda_2}{2}n^2_y +\lambda_3(l_xl_z-k_xk_z);\\
&B_1: \, -\frac{\lambda_1}{2}n^2_x-\frac{\lambda_2}{2}n^2_y \pm\lambda_4(l_xk_y+k_xl_y);\\
&B_2: \, -\frac{\lambda_1}{2}n^2_x-\frac{\lambda_2}{2}n^2_y+ \lambda_3(l_xl_z-k_xk_z)+\\ &\, \pm\lambda_4(l_xk_y+k_xl_y);\\
&C_1: \, \frac{\lambda_1}{2}l^2_x+\frac{\lambda_2}{2}l^2_y +2\lambda_3l_xl_z;\\
&C_2: \, -\frac{\lambda_1}{2}n^2_x-\frac{\lambda_2}{2}n^2_y -\lambda_3n_xn_z;\\
&D: \, \frac{\lambda_1}{2}l^2_x+\frac{\lambda_2}{2}l^2_y\mp2\lambda_4l_xl_y;\\
&E: \, -\frac{\lambda_1}{2}n^2_x-\frac{\lambda_2}{2}n^2_y \pm\lambda_4n_xn_y;\\
&D^*: \, \frac{\lambda_1}{2}l^2_x+\frac{\lambda_2}{2}l^2_y;\\
&E^*: \, -\frac{\lambda_1}{2}n^2_x-\frac{\lambda_2}{2}n^2_y.
\end{split}\end{equation*}
Here for the planar structures we introduced the normal vector to the spin plane ${{\bf n}=[{\bf l}{\bf k}]}$
and used the relation
\[l_{\alpha}l_{\beta}+k_{\alpha}k_{\beta}+n_{\alpha}n_{\beta}= \delta_{\alpha\beta}.\]

Electric polarization of relativistic nature is due to the following invariants:
\begin{equation}\label{Ep}\begin{split}
&-d_1E_x(\mu_x\nu^*_z+\mu^*_x\nu_z+\mu_z\nu^*_x+\mu^*_z\nu_x)\\
&-d_2E_y(\mu_y\nu^*_z+\mu^*_y\nu_z+\mu_z\nu^*_y+\mu^*_z\nu_y)\\
&-d_3E_z(\mu_z\nu^*_z+\mu^*_z\nu_z).
\end{split}\end{equation}
Accordingly, in five phases of twelve components of polarization $P_x,P_y,P_z$ occur:
\begin{equation*}\label{EFA1}\begin{split}
&A_1)\, d_1(l_xk_z+k_xl_z),d_2(l_yk_z+k_yl_z),d_3l_zk_z;\\
&A_2)\, d_1(l_xk_z+k_zl_x),d_2(l_yk_z+k_yl_z),d_3l_zk_z;\\
&B_1)\, d_1(l_xl_z-k_xk_z),d_2(l_yl_z-k_yk_z),d_3(l^2_z-k^2_z);\\
&D^*)\, 2d_1l_xl_z,2d_2l_yl_z,P_0+d_3l^2_z;\\
&E^*)\, d_1(l_xl_z+k_xk_z),d_2(l_yl_z+k_yk_z),P_0-d_3n^2_z.
\end{split}\end{equation*}
Here for the phases $D^*$ и $E^*$ we added the already mentioned polarization of exchange nature $P_0$.
Note that $P_0$ and the relativistic contribution to the polarization, as well as the "anomalous" \ terms of anisotropy $(\lambda_3,\lambda_4)$ are given for the one of four possible domains, differed by the sign of ${\boldsymbol{\nu}}$ and the replacement ${\boldsymbol{\mu}\leftrightarrows\boldsymbol{\nu}}$.

Thus, if we exclude the collinear structure $D^*$, there are four candidates for the phase, observed in $LiCu_2O_2$ at the temperature ${T<23,2\,K}$.

Due to the low symmetry of the crystal, there are too many invariants caused by the relativistic effects of magnetostriction:
$\eta_{xx}u_{xx},$ $\eta_{xx}u_{yy},$ $\eta_{xx}u_{zz},$ $\eta_{yy}u_{xx},$ $\eta_{yy}u_{yy},$ $\eta_{yy}u_{zz};$
$\eta_{zz}u_{xx},$ $\eta_{zz}u_{yy},$ $\eta_{zz}u_{zz};$
$\eta_{xy}u_{xy},$ $\eta_{yz}u_{yz},$ $\eta_{zx}u_{zx};$
$r_{xx}u_{xz},$ $r_{yy}u_{xz},$ $r_{zz}u_{xz},$ $r_{xz}u_{xx},$ $r_{xz}u_{yy},$ $r_{xz}u_{zz},$ $r_{xy}u_{yz},$ $r_{yz}u_{xy};$ $s_{xy}u_{xx},$ $s_{xy}u_{yy},$ $s_{xy}u_{zz},$
$s_{xx}u_{xy},$ $s_{yy}u_{xy},$ $s_{zz}u_{xy},$ $s_{yz}u_{xz},$ $s_{xz}u_{zy};$
here, the following forms are introduced for brevity
\[\eta_{ik}=\mu_i\mu^*_k+\mu_k\mu^*_i+\nu_i\nu^*_k+\nu_k\nu^*_i,\]
\[r_{ik}=\mu_i\mu^*_k+\mu_k\mu^*_i-\nu_i\nu^*_k-\nu_k\nu^*_i,\]
\[s_{ik}=\mu_i\nu^*_k+\mu_k\nu^*_i-\mu^*_i\nu_k-\mu^*_k\nu_i,\]
therefore the use of strain measurements for the task of structure selection is difficult.

{\bf{Spin-orbit splitting of phase transition.}}
The above regular accounting for relativistic effects under the perturbation theory may become inapplicable in the vicinity of the transition, where their contribution necessarily becomes comparable to the exchange quadratic term of the Dzyaloshinskii-Landau expansion $\tau\eta^2$. In this case, an intermediate magnetic phase may appear in a small vicinity of the transition in the cases, when active representation of the exchange approximation is not one-dimensional. Let us find out the features of this phase, suggesting the Lifshitz instability is weak.

Let us select the real and imaginary parts in the order parameter ${\eta\boldsymbol{\mu}={\bf a}+i{\bf b}},$
${\eta\boldsymbol{\nu}={\bf c}+i{\bf d}}$, without imposing any restrictions on the real-valued vectors ${\bf{a}},$ ${\bf{b}},$ ${\bf{c}},$ ${\bf{d}}$. In these variables the quadratic part of the Dzyaloshinskii-Landau expansion has the form
\begin{equation}\label{Fanabcd}\begin{split}
&(\tau+\lambda_1)(a_x^2+b_x^2+c_x^2+d_x^2)+\\ &+(\tau+\lambda_2)(a_y^2+b_y^2+c_y^2+d_y^2)+\\
&+\tau(a_z^2+b_z^2+c_z^2+d_z^2)+\\
&+2\lambda_3(a_xa_z+b_xb_z-c_xc_z-d_xd_z)+\\
&+2\lambda_4(a_xd_y-b_xc_y+d_xa_y-c_xb_y).
\end{split}\end{equation}

In the immediate vicinity of the transition we can not neglect the Lifshitz invariant (\ref{L}), and should also
consider suppressive instability quadratic in the gradients exchange invariant
\begin{equation}\label{Labcd}\begin{split}
{\cal{L}}({\bf{a}}{\bf{c}}'-{\bf{c}}{\bf{a}}'+
{\bf{b}}{\bf{d}}'-{\bf{d}}{\bf{b}}')+\\+
g({\bf{a}}'^2+{\bf{b}}'^2+{\bf{c}}'^2+{\bf{d}}'^2),
\end{split}\end{equation}
where prime means differentiation with respect to coordinate $x$.

In the general case this is a rather complicated task. However, in LiCu$_2$O$_2$, according to the experimental data,\cite{Svistov_2012} if the parameter $\lambda_1$ is negative and substantially exceeds the value of the remaining anisotropy constants, then the analysis of the transition is much easier. The transition then should occur at ${\tau\approx-\lambda_1}$
and we may neglect $y$ and $z$ components of the vectors in the vicinity of $T_{c1}$. As a result, the quadratic term in the Dzyaloshinskii-Landau expansion in the order parameter has the form
\begin{equation}\label{DLabcd}\begin{split}
F_2=\tilde{\tau}\eta^2+
{\cal{L}}(ac'-ca'+bd'-db')+\\+
g(a'^2+b'^2+c'^2+d'^2),
\end{split}\end{equation}
where ${\eta^2=a^2+b^2+c^2+d^2}$, the same for all vectors index $x$ has been omitted.

Let us turn to the Fourier components ${a=a_ke^{ikx}+a^*_ke^{-ikx},\,...,\, k>0}.$
For the contribution of single harmonic we have
\begin{equation}\label{Tabcd}\begin{split}
F_2=(\tilde{\tau}+gk^2)(a_k^*a_k+b_k^*b_k+c_k^*c_k+d_k^*d_k)+\\+ ik{\cal{L}}(a_k^*c_k-c_k^*a_k+b_k^*d_k-d_kb_k^*).
\end{split}\end{equation}
For given magnitude ${a_k^*a_k+b_k^*b_k+c_k^*c_k+d_k^*d_k}$ and wavevector this contribution is minimal if
\begin{equation}\label{cd}c_k=ia_k, \, d_k=ib_k, \end{equation}
then
\begin{equation}\label{Tk}
F_2=2(\tilde{\tau}-{\cal{L}}k+gk^2)(a_k^*a_k+b_k^*b_k).
\end{equation}

The phase transition occurs when the temperature decreases and the minimal value of the coefficient in (\ref{Tk})
changes sign at ${k=k_0={\cal{L}}/2g}$. The relative phase and amplitude of the components $a_k$ and $b_k$
are determined by the minimum condition of the function
\begin{equation}\label{Babcd}\begin{split}
&{\cal{B}}=4\beta_1(ac+bd)^2+\\
&\beta_2(a^2+b^2-c^2-d^2)^2+4\beta_3(ad-bc)^2+\\
&\beta_4[(a^2-b^2+c^2-d^2)^2+4(ab+cd)^2]+\\
&\beta_5[(a^2-b^2-c^2+d^2)^2+4(ab-cd)^2]+\\
&4\beta_6(a^2c^2+b^2d^2+a^2d^2+b^2c^2)
\end{split}\end{equation}
at the fixed magnitude ${\eta^2=2(a_k^*a_k+b_k^*b_k)}$ and fulfillment the condition (\ref{cd}).

Turning to the Fourier components in the expression (\ref{Babcd}) and integrating over the volume we get
\begin{equation*}\begin{split}
&8(\beta_1+\beta_2)(a^2+b^2)(a^{*2}+b^{*2})-16\beta_3(a^*b-ab^*)^2\\
&+16\beta_4(a^2a^{*2}+b^2b^{*2}+a^2b^{*2}+a^{*2}b^2)\\
&+8(\beta_5+\beta_6)[(a^2-b^2)(a^{*2}-b^{*2})+4aa^*bb^*],
\end{split}\end{equation*}
here the indices $k$ have been omitted.

Introducing parametrization
\begin{equation}\label{abk}
a_k=\frac{\eta}{2}e^{-i\varphi}s_\phi, \,b_k=\frac{\eta}{2}c_{\phi}e^{i(\delta-\varphi)},
\end{equation}
we get
\begin{equation}\label{Bab}
{\cal{B}}={\cal{B}}_0+{\cal{A}}s^2_{2\phi}s^2_\delta,
\end{equation}
where
\begin{equation}\label{A}\begin{split}
&{\cal{B}}_0=\frac{1}{2}(\beta_1+\beta_2+2\beta_4+\beta_5+\beta_6),\\
&{\cal{A}}=-\frac{1}{2}(\beta_1+\beta_2+2\beta_3+2\beta_4+\beta_5+\beta_6).
\end{split}\end{equation}

If ${{\cal{A}}<0}$ then the minimum ${\cal{B}}$ occurs at ${\phi=\pi/4}$, ${\delta=\pi/2}$. If ${{\cal{A}}>0}$ then
there are two solutions ${1)\,\phi=0}$ (or ${\phi=\pi/2}$), and the value of $\delta$ is not relevant, and 2) ${\delta=0,\pi}$. Thus, the following solutions are the candidates for the intermediate phase
\begin{flalign*}
I_1:\mu_x&=\pm{i}\nu_x=\frac{e^{-i(kx-\varphi)}}{\sqrt{2}}\\
I_2:\mu_x&=\cos(kx-\varphi), \, \nu_x=\pm\sin(kx-\varphi);\\
I_3:\mu_x&=\cos(kx-\varphi), \, \nu_x=\pm{i}\sin(kx-\varphi).
\end{flalign*}

Let us note that there are no associated solutions transforming into each other under the transformation (\ref{inunu}) here, since, as noted above, the Lifshitz invariant (\ref{L}) breaks this random symmetry.
One of these phases has to be realized at the transition point $T_ {c1}$. Note that only in the phase $I_2$ electric polarization of exchange nature occurs.

There are two possible scenarios with decreasing temperature: either a first-order transition into one of the phases considered above at the temperature $T_{c2}$ or, if the low-temperature phase is noncollinear, then a first-order transition into another intermediate collinear phase $C_1,D$ or $D^*$ is possible.


\begin{thebibliography}{37}
\bibitem{Redundant} Since redundant solutions may occur at deriving (\ref{AE}), all the solutions must be checked by direct substitution into the original equations (\ref{mu},\ref{nu}).
\bibitem{Contact_field_note} We tried to fit the experimental spectra in the model where the second pair of copper ions induces the dominant contact field. This assumption does not change the conclusions of our paper.
\end{thebibliography}
\end{document}